\documentclass[prb,aps,twocolumn,amsmath,amssymb,floatfix,citeautoscript]{revtex4}
\usepackage{amssymb}
\usepackage{amsmath}
\usepackage{graphicx}
\usepackage{bm}
\usepackage{dcolumn}

\newcommand{\beq}{\begin{equation}}
\newcommand{\enq}{\end{equation}}
\newcommand{\kB}{k_\mathrm{B}}

\begin{document}

\title{Local and chain dynamics in miscible polymer blends: A Monte
  Carlo simulation study} 
\author{Jutta Luettmer-Strathmann}
\email[]{jutta@physics.uakron.edu}
\affiliation{
Department of Physics and Department of Chemistry, The University of
Akron, Akron, Ohio 44325-4001} 
\author{Manjeera Mantina}
\altaffiliation{
present address:
Department of Materials Science and Engineering, The Pennsylvania
State University, University Park, PA 16802; email: mantina@psu.edu  
}
\affiliation{
Department of Computer Science and Department of Chemistry, The
University of Akron, Akron, Ohio 44325}
\date{\today}
\begin{abstract}
Local chain structure and local environment play an important role in
the dynamics of polymer chains in miscible blends.  In general, the
friction coefficients that describe the segmental dynamics of the two
components in a blend differ from each other and from those of the
pure melts. In this work, we investigate polymer blend dynamics with
Monte Carlo simulations of a generalized bond-fluctuation model, where
differences in the interaction energies between non-bonded nearest
neighbors distinguish the two components of a blend. Simulations
employing only local moves and respecting a non-bond crossing
condition were carried out for blends with a range of compositions,
densities, and chain lengths. The blends investigated here have
long-chain dynamics in the crossover region between Rouse and
entangled behavior. 
In order to investigate the scaling of the self-diffusion
coefficients, characteristic chain lengths $N_\mathrm{c}$ are
calculated from the packing length of the chains. 
These are combined with a local mobility $\mu$ determined from the
acceptance rate and the effective bond length to yield characteristic
self-diffusion coefficients $D_\mathrm{c}=\mu/N_\mathrm{c}$. 
We find that the data for both melts and blends collapse onto a common
line in a graph of reduced diffusion coefficients $D/D_\mathrm{c}$
as a function of reduced chain length $N/N_\mathrm{c}$. 
The composition dependence of dynamic properties is investigated in 
detail for melts and blends with chains of length twenty at three
different densities. 
For these blends, we calculate friction coefficients from the
local mobilities and consider their composition and pressure
dependence. 
The friction coefficients determined in this way 
show many of the characteristics observed in experiments on miscible
blends.  
\end{abstract}

\maketitle

\section{Introduction}

Processes on different length scales affect dynamic properties of 
polymer melts and blends \cite{do86,gr94c,de79,fe80}. 
The dynamics of miscible blends have been the subject of a large
number of recent investigations with experimental (see, for example, 
Refs. \hspace{0em} 
\onlinecite{ro91b,ro92,ch94,he03,ch94b,al94,ki95,pa01,%
mi01,ha03,lu03d,ha04,lu04,al02,do00,ge05}), 
simulation (cf.\ Refs. \hspace{0em} 
\onlinecite{ge05,ko97c,ka03b,do00,bu02,fa04,ne05}), 
and theoretical methods
(cf.\ Refs. \hspace{0em}
\onlinecite{ze94,ka95b,ku96,ka99,lo00,he03,le03,ka03c,ch94,%
ro91b,ro92,ng04,lu05}). 
Monomeric friction coefficients, which are inversely proportional to
the mobility of short chain segments, are a convenient way to describe
the effect of local dynamic properties on global dynamic
properties such as the viscosity \cite{fe80}. 

Since blending changes the local environment of the chain segments of
a polymer it has a strong effect on the local dynamics of the chains. 
 From experimental \cite{ro91b,ro92,ch94,ch94b,al94,ki95,do00,pa01,%
mi01,ha03,lu03d,ha04,lu04,al02,he03}
and simulation work  \cite{ko97c,ka03b,do00,bu02,fa04,ne05,ge05}
it is found that the local dynamics of the two blend 
components differ from each other and the pure melts. 
The addition of slow (high friction coefficient) component to a blend
is found to increase the friction coefficients of both components and,
conversely, the addition of fast component is found to speed up both
components. 
The effects of blending are most pronounced near the glass
transition, however, they are observed even at high temperatures 
\cite{do00,mi01,ha03,lu03d} and
in blends where one of the components is dilute \cite{lu03d,lu04,ha04}.

Several recently developed models for miscible polymer blends 
relate differences in the component dynamics to local variations in
the glass transition temperature induced by local variations 
in blend composition \cite{ze94,ka95b,ku96,ka99,lo00,he03,le03,ka03c}, 
while others consider  ``intrinsic'' effects, 
due to differences in the chain structure of the two components, in
addition to local density and composition variations
\cite{ch94,ro91b,ro92,ng04,lu05}.  
Unfortunately, it is still difficult, in general, to predict the local
friction coefficients in a blend from those of the melts.

Self diffusion coefficients give information about the chain 
dynamics of polymer melts and blends.  
The Rouse model prediction for the self-diffusion 
coefficient of a polymer chain may be written as \cite{do86,gr94c}. 
\beq\label{DRouse}
D_\mathrm{R} = \frac{\kB T}{\zeta N} = \frac{\mu}{N} ,
\enq 
where $T$ is the temperature, $\kB$ is Boltzmann's constant, $N$ is
the chain length, $\zeta$ is the so-called monomeric friction
coefficient, and $\mu=\kB T/\zeta$ is the corresponding mobility. 
The long-time dynamics of long polymer chains are dominated by
entanglement effects, which are not part of the Rouse model. 
The reptation model \cite{do86,gr94c,de79} describes the entanglements
of a chain with other chains in terms of a tube,  
which restricts the motion of the chain perpendicular to the tube. 
If the average number of monomers between entanglements is denoted by 
$N_\mathrm{e}$, then the reptation prediction for the self-diffusion
coefficient takes the form
\begin{equation}\label{Drep}
D_\mathrm{G} = 
\frac{\kB T N_\mathrm{e}}{3N^2\zeta} = 
\frac{1}{3}\frac{N_\mathrm{e}}{N} D_\mathrm{R}. 
\end{equation}
The entanglement length of a polymer melt may be determined directly
from experimental data on the plateau modulus \cite{do86,gr81}. 
For viscoelastic properties, 
it differs by a constant prefactor from the characteristic chain
length $N_\mathrm{c}$, that separates short chain (unentangled) 
behavior from long chain (entangled) behavior \cite{gr81}.
Experimental, theoretical and simulation work on polymer melts
suggests that the tube diameter of the reptation theories is
proportional to the so-called packing length
\cite{li87,ka87,fe99,ev04,su05,mi05},   
which may be defined as \cite{fe99,ev04}
\beq\label{packing}
p=\frac{N}{\rho R_e^2},
\enq
where $\rho$ is the monomer density and $R_e^2$ is the average squared
end-to-end vector of the chains.
Hence, the entanglement length is expected to be proportional to the
chain length corresponding to $p$, which yields \cite{mi05}
\beq\label{Ncfromp}
N_\mathrm{c} \sim p^3\rho .
\enq

The transition between the unentangled and entangled regimes
is not sharp and simulation data are often found to be in
the crossover region between unentangled and reptation behavior
\cite{bi97,mu00c,ta00,kr01,pa04b,le05}. 
In the following, it is convenient to rescale the chain length
$N$ by the characteristic chain length $N_\mathrm{c}\propto
N_\mathrm{e}$ and the  
self-diffusion coefficient $D$ by a characteristic diffusion
coefficient $D_\mathrm{c}$ defined as \cite{mu00c}, 
\beq\label{Dc}
D_\mathrm{c}=\frac{\mu}{N_\mathrm{c}} .
\enq
In this way, the 
predictions for the self-diffusion coefficient may be summarized as 
\begin{eqnarray}\label{DbyDc}
\frac{D}{D_\mathrm{c}} &=& \left\{ 
\begin{array}{lcl}
\left({N}/{N_\mathrm{c}}\right)^{-1} & 
\textrm{ for } & N\ll N_\mathrm{c} \\
\left({N}/{N_\mathrm{c}}\right)^{-2} 
& \textrm{ for } & N\gg N_\mathrm{c} 
\end{array}
\right. 
\end{eqnarray}
so that, at the chain length $N_\mathrm{c}$, the extrapolations from 
both power laws yield the same value, $D/D_\mathrm{c}=1$. 
For long but not completely entangled chains, 
Hess \cite{he86,he88} has proposed a crossover equation for the 
self-diffusion coefficient, which may be written in our notation as 
\begin{equation}\label{DHess}
\frac{D}{D_\mathrm{c}} = 
\frac{\displaystyle \left({N}/{N_\mathrm{c}}\right)^{-1} }
{\displaystyle 1 + {N}/{N_\mathrm{c}} } .
\end{equation}
This expression has been found to give a good representation of
data for self-diffusion coefficients of the bond-fluctuation model
\cite{pa91}. However, the resulting values for the characteristic
chain length $N_\mathrm{c}$ are considerably smaller than those
obtained by superimposing simulation data with experimental data that
extends deeply into the entangled regime \cite{ta00}.

Simulation work on polymer blends has been carried out with atomistic
\cite{do00,bu02,fa04,ne05,ge05} and coarse grained models
\cite{ko97c,ka03b}. 
Molecular dynamics simulations of atomistic models give access to
the chain structure and dynamics of realistic polymers and 
allow a detailed investigation of the environments of the chain
segments \cite{ne05}.
Simulations of coarse grained models, on the other hand, can be used
to isolate particular effects such as isotope effects \cite{ko97c} and
differences in chain stiffness \cite{ka03b}. 

In this work, we investigate chain and local dynamics of miscible
polymer blends with the aid of Monte Carlo simulations of a lattice
model. We are interested in energetic effects and represent the two
components of a blend by chains that differ only in interaction
energies. 
The model is a modification of Shaffer's bond-fluctuation model for
athermal melts \cite{sh94,sh95,pa96} and is introduced in Section
\ref{model}.   
Section \ref{model} also provides some details about the Monte Carlo
simulations. In Section \ref{evaluation} we discuss the evaluation of
the Monte Carlo simulations and describe the construction of 
scaling variables for the self-diffusion coefficients. 
Results of our simulations are presented in Section \ref{results}
and discussed in Section \ref{discussion}. 

\section{Blend model for Monte Carlo simulation}\label{model}

Shaffer's bond fluctuation model \cite{sh94,sh95,pa96} is a lattice
model for polymer chains, where the monomers occupy sites of a simple
cubic lattice.  
In the following, we will take the size of the unit cell, i.e.\ the
lattice constant $a$,  as the unit for length.  
The monomers are connected by bonds of three possible lengths, namely 
1, $\sqrt{2}$, and $\sqrt{3}$, corresponding to the 
sides, face diagonals, and body diagonals of the unit cell of the
lattice. 
Monte Carlo simulations of the model employ only local moves, where an 
attempt is made to displace a randomly chosen monomer by one
lattice site along any of the three coordinate directions. 
One attempted elementary move per monomer in the system is called one 
Monte Carlo step (MCs) and will be our unit of time. 
Shaffer considered two versions of this model. In the first version,
bonds are allowed to cross each other with the result that the chains
do not entangle; in the second, bond crossings are prohibited and
entanglement effects become apparent. We adopt the second approach and
prohibit bond crossings in this work. 

\begin{figure}
\includegraphics*[width=2.3in]{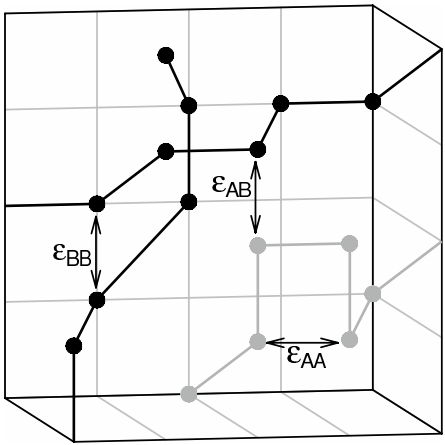}
\caption{\label{fig1}
Illustration of the generalization of Shaffer's bond fluctuation model
to polymer blends. 
The circles connected by heavy lines represent polymer chains in a blend, 
gray for component A and black for component B. Allowed bond lengths
are 1, $\sqrt{2}$ and $\sqrt{3}$ corresponding to the sides, face
diagonals, and body diagonals of the unit cell of the simple cubic 
lattice.  
The arrows show examples of nearest
neighbor interactions between like ($\epsilon_{AA}$ and
$\epsilon_{BB}$) and unlike ($\epsilon_{AB}$) monomers.}
\end{figure}

Shaffer's model describes athermal melts; the monomers interact only
through hard core repulsion, which is enforced by prohibiting double
occupation of lattice sites. \cite{sh94} 
In this work, we are interested in  blends of polymers that have
identical bond structures but differ in their monomer-monomer
interactions.  
To this end, we add attractive interactions between 
non-bonded monomers occupying nearest-neighbor sites to the model, see
Fig.~\ref{fig1}. 
For a binary blend of chains
of type A and B, the energy parameters $\epsilon_{AA}$,
$\epsilon_{BB}$, and $\epsilon_{AB}$, describe interactions between
monomers of type A, type B, and mixed interactions, respectively. 
The total internal energy of the system is given by
\beq\label{E}
E = N_{AA}\epsilon_{AA} + N_{BB}\epsilon_{BB} + 
N_{AB}\epsilon_{AB} ,
\enq
where $N_{ij}$, $i,j\in \{ A,B\}$, denotes the number of nearest
neighbor contacts between monomers of type $i$ and $j$.  
In this work, we choose $\epsilon_{AA}=-1\epsilon$, 
$\epsilon_{BB}=-2\epsilon$, and $\epsilon_{AB}=-2\epsilon$, where
$\epsilon$ is the unit of energy. 
The large difference in the interaction energies $\epsilon_{AA}$ and
$\epsilon_{BB}$ makes differences between A and B chains readily
observable in simulations. The value of $\epsilon_{AB}$ corresponds to
very attractive interactions between unlike monomers and insures the
miscibility of the blends. 
From the temperature $T$, the unit of energy, $\epsilon$, and
Boltzmann's constant, $\kB$, 
a dimensionless temperature $T^*$ and its inverse $\beta$ are 
defined as $T^* = \kB T/\epsilon$ and $\beta = 1/T^*$. 
In this work, we present blend simulation results for the
fixed temperature $T^* = 10$ corresponding to $\beta = 0.1$. 

For a cubic lattice with $L$ lattice sites on the side, the monomer
density $\rho$ is defined as 
\beq\label{rho}
\rho = \frac{ N_{pA}N_A + N_{pB}N_B }{L^3}, 
\enq
where $N_{pA}$ and $N_{pB}$ are the number of chains, while
$N_A$ and $N_B$ are the chain lengths for chains of type $A$ and $B$,
respectively. 
The mass fraction $c_A$ of component $A$ in a blend is given by
\beq
c_A = \frac{N_{pA}N_A}{N_{pA}N_A + N_{pB}N_B } . 
\enq
In this work, we consider only monodisperse melts and
blends, i.e.\ $N_A = N_B \equiv N$, so that the mass fraction is equal
to the mole fraction. 

Shaffer established that a monomer density of $\rho=0.5 a^{-3}$
corresponds to a dense melt for the athermal system \cite{sh94}. 
We performed Monte Carlo simulations for the three  monomer
densities $\rho = 0.5 a^{-3}$, $0.6 a^{-3}$, and $0.7 a^{-3}$ at
$\beta = 0.1$. For melts and 
50/50 blends, we 
considered nine chain lengths between $N=5$ and $N=80$. 
For intermediate concentrations, we focused on chains of length
$N=20$. Simulations were performed on a lattice of size $L=20$ with 
periodic boundary conditions applied along the three coordinate
directions. In order to test for finite size effects, we performed
simulations for chains of length $N=80$ in a 50/50 blend on a lattice
of size $L=40$ and found no significant differences in the results. 

Initial configurations were created by randomly placing dimers on the
lattice and repeatedly reassembling shorter chains into longer chains,
where the no bond-crossing condition was enforced from the start. 
These initial configurations were equilibrated at the inverse
temperature $\beta = 0.1$ before a trajectory of the simulation
consisting of 10,000 configurations separated by a fixed number of
Monte Carlo steps, $t_r$, was recorded.  The number of Monte 
Carlo steps during production varied between $10^6$ for the shortest
chains and $10^8$ for the longest chains. Typical simulation times for
chains of length $N=20$ are $10^7$ Monte Carlo steps. In order to
improve the statistics for blends, where the mass fraction of one of
the components is very small (5\%), simulation times were extended to
at least $5\times 10^7$. 
In all simulations, the equilibration time was at least 10\% of the
production time and was sufficiently long for the chains to travel
a distance corresponding to multiple times the radius of gyration. 
During the simulations, the acceptance rates $A_{\text{rate}}$ 
for elementary moves are monitored separately for each of the
components.  

\section{Evaluation of simulation results}\label{evaluation}

The Monte Carlo simulations are evaluated to yield static and dynamic
quantities. The average radius of gyration, $R_{g,A}$, of chains of
type A, for example, is calculated from  
\beq\label{Rg}
R_{g,A}^2 = \frac{1}{N_{pA}}\sum_{k=1}^{N_{pA}}
\frac{1}{N^2}\sum_{i<j}\langle (\mathbf{r}_{i,k}-\mathbf{r}_{j,k})^2  
\rangle , 
\enq
where $\mathbf{r}_{i,k}$ is the position of monomer $i$ on chain $k$
of component A, 
and where the angular brackets indicate the average over 
configurations in the trajectory. 
The calculation for the radius of gyration of chains of type B
proceeds in the same way. In order to simplify notation, we drop 
the subscripts indicating the type of chain whenever the
calculation is identical for both components in the blend and there is
no danger of confusion. 
Error estimates for static quantities are obtained from block
averaging \cite{ne99}, where we divide the trajectories into ten equal
blocks. 

In melts and miscible blends, polymer chains of sufficient length are
expected to obey Gaussian statistics. For such chains, the average
squared end-to-end distance $R_e^2$ is proportional to the number of
bonds, $R_e^2=b^2(N-1)$, where $b$ is the so-called effective bond length
\cite{do86}. Since the squared averages of the radius of gyration 
and end-to-end distance for dense systems are related through
$R_e^2=6R_g^2$, we calculate the effective bond length from 
\beq\label{b2}
b^2 = 6R_g^2/(N-1). 
\enq
For Gaussian chains, the packing length defined in Eq.~(\ref{packing})
may be expressed as  
$p = \left( \rho b^2 \right)^{-1}$,
where $\rho$ is the monomer density defined in Eq.~(\ref{rho}). 
This allows us to calculate an estimate for the characteristic chain
length $N_\mathrm{c}$ from Eq.~(\ref{Ncfromp}) 
\beq\label{Ncsims}
N_\mathrm{c} =  C_c \left( b^3 \rho \right )^{-2} ,
\enq 
where the constant of proportionality, $C_c$, is determined from a
comparison with the crossover equation (\ref{DHess}) of Hess
\cite{he86,he88}.

In this work, we present results for two mean-squared displacement
functions. 
The mean squared displacement of the center of mass is obtained from  
\beq\label{gd}
g_{d}(t) = \frac{1}{N_{p}}\sum_{k=1}^{N_{p}}
\langle
(\mathbf{r}_{\mathrm{cm},k}(t)-\mathbf{r}_{\mathrm{cm},k}(0))^2
\rangle ,  
\enq
where $\mathbf{r}_{\mathrm{cm},k}$ is the position of the center of
mass of chain $k$ and where
the angular brackets indicate the average over 
configurations that are $t$ Monte Carlo steps apart. 
Similarly, the mean squared displacement of the central monomer is
obtained from 
\beq\label{g1}
g_1(t) = \frac{1}{N_{p}}\sum_{k=1}^{N_{p}}
\langle
(\mathbf{r}_{N/2,k}(t)-\mathbf{r}_{N/2,k}(0))^2
\rangle 
\enq
where $\mathbf{r}_{N/2,k}$ represents the position of the central monomer
of the chain (the results are averaged over both innermost monomers for
chains with an even number of monomers).

\begin{figure}
\includegraphics*[width=3in]{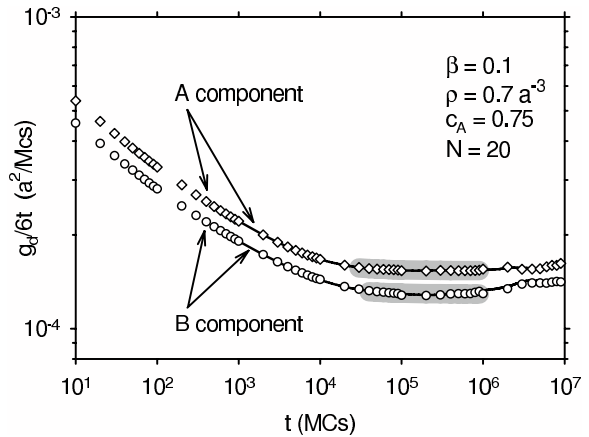}
\caption{\label{figgd}
The function $g_d/6t$, where $g_d$ is the 
mean squared displacement and  $t$ is the time, as a function of 
time for an A-rich blend with mass fraction $c_A=0.75$, monomer 
density $\rho = 0.7 a^{-3}$, inverse temperature $\beta = 0.1$, and chain
length $N=20$.  
The symbols indicate results for the two components, A 
(diamonds) and B (circles), obtained with the block algorithm during
the simulation. The black lines represent results
from the evaluation of the trajectory, which is limited to times
larger than the recording interval $t_r = 10^3$ MCs and agrees well
with the results from the block algorithm.  
The gray areas show the ranges of $g_d/6t$ data that were used 
to determine the self-diffusion coefficients of the two components as
explained in the text. 
}
\end{figure}

The self-diffusion coefficients of the components are determined from
the long-time limit of mean-squared displacement $g_d$ of the center
of mass  
\beq\label{D}
D = \lim_{t\rightarrow\infty} \frac{g_{d}(t)}{6t} .
\enq
In Fig.~\ref{figgd} we present simulation results for $g_{d}/6t$ 
for both components of an A-rich blend ($c_A=0.75$) of chains of length
$N=20$ at a monomer density of $\rho=0.7$ and an inverse temperature of
$\beta=0.1$. 
The results in Fig. \ref{figgd} were calculated by two different
methods. 
During the simulations, a block algorithm adapted from Frenkel and Smit
\cite{fr96b} was used to determine mean squared displacements. 
The algorithm yields values of the displacement functions 
at intervals that increase with increasing time and, thus, gives access
to a large time range.  These values are indicated by open symbols in
Fig. \ref{figgd}. 
We also calculated displacement functions from the trajectories at the
time intervals $t_r$ for times up to half the total run time. 
These results are represented by solid lines in Fig.~\ref{figgd}. 
The graphs show that the values for $g_d/6t$ obtained by the two
methods agree well with each other. The block algorithm is very
efficient and was used to determine most of the self-diffusion
coefficients presented in this work. 
The results for $g_d/6t$ decrease with time at short times before they
level off to a constant value and, finally, become irregular at very
long times. 
The decrease of $g_d/6t$ at short times indicates the 
subdiffusive behavior expected for shorter chains \cite{kr01,pa02b}. 
For very long times, the results for the displacement functions are
irregular since they represent averages over few configurations so
that individual events, like the release of a chain from
entanglements, can alter the shape of the functions \cite{mc05}. 
The $g_d/6t$ values presented in Fig.~\ref{figgd} for the B component
are more noisy than those for the A component since the blend is rich
in A and contains fewer chains of type B.
Self-diffusion coefficients are determined from the data in the time
range where $g_d/6t$ is nearly constant; the range starts 
when $g_d/6t$ closely approaches its asymptotic plateau
and excludes the largest times, where the results for $g_d/6t$ become
irregular.  
In Fig.~\ref{figgd} we show as gray-shaded areas the data ranges that
were used to determine self-diffusion coefficients for the blend
components. 
For each component, we calculate the average of the values of 
$g_d/6t$ in this range, and we also fit a function of the form
$f(t)=D+\text{const}/6t$ to the data. 
When the times in the fit range are sufficiently long, the second term
in $f(t)$ is very small and the values of the diffusion coefficients
obtained by the two methods agree with each other within their
statistical uncertainty. In this work, we took care to have simulation
runs of sufficient length to allow for a consistent determination of
the self-diffusion coefficients.

In order to obtain estimates for the local friction coefficients from
our simulations, we combine measurements of the effective 
bond length with results for the acceptance rates of elementary moves.
According to the Rouse model, the segmental mobility $\mu$ is related
to the shortest Rouse relaxation time through \cite{do86}
\beq\label{tauN}
\mu = \frac{1}{3\pi^2}\frac{b^2}{\tau_N},
\enq
where $b$ is the effective bond length and $\tau_N$ is 
the relaxation time of a single segment. 
Since $\tau_N$ is expected to be inversely proportional to the
acceptance rate, which measures the probability that a monomer  
is able to complete an attempted move to a nearest-neighbor site, we
set in this work 
\beq\label{musim}
\mu = C_\mu b^2 A_{\text{rate}},
\enq
where $A_\text{rate}$ is the acceptance rate. $C_\mu$ is a constant of
proportionality that is determined from the requirement that
$D/D_\mathrm{c}=1$ for $N/N_\mathrm{c}=1$, see Eqs.~(\ref{Dc}) and 
(\ref{DbyDc}). 
The local friction coefficients, finally, are determined from 
\beq\label{zetasim}
\zeta = \kB T/\mu .
\enq

\section{Results}\label{results}

\begin{figure}
\includegraphics*[width=3in]{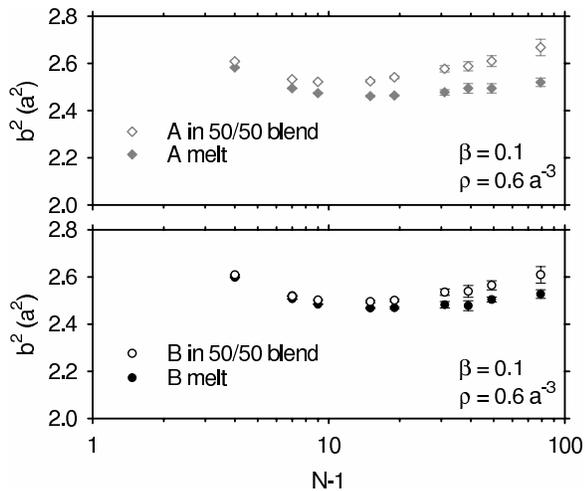}
\caption{\label{figb2}
Simulation results for the squared effective bond lengths, $b^2 =
6R_g^2/(N-1)$, of polymer chains as a function of the number of
bonds, $N-1$, for a 50/50 blend and for the A and B melts at the
temperature and monomer density indicated in the figure. 
The upper and lower panels show the results for chains of type A and
B, respectively. 
A comparison of the chain dimensions in the blends
(open symbols) with those in the melts (filled symbols) shows that
both types of chains are expanded in the blend.
}
\end{figure}

The effective bond length, which is related to the radius of gyration
through Eq.~(\ref{b2}) depends on the structure as well as the
environment of the chains. 
In Fig.~\ref{figb2} we present simulation results for the squared
effective bond length as a function of chain length for  A and B
chains in the melt and in a blend with mass fraction $c_A=0.5$. 
We note, first of all, that for both melts and blends the bond length
approaches a constant value for long chain lengths, as expected 
for Gaussian chains, see Eq.~(\ref{b2}). 
Furthermore, the effective bond length of both A and
B chains is larger in the blend than in the melt. 
At all densities, the chain expansion increases with increasing
dilution, i.e. chains are most expanded when they are surrounded
by chains of the other component. 
This is due to the relatively large attractive
interaction energy associated with contacts between unlike monomers; 
the mixed interaction with energy $\epsilon_{AB}=-2\epsilon$ is twice as
attractive as the interaction between A segments,
$\epsilon_{AA}=-1\epsilon$, so that A monomers seek out B monomers
as neighbors, which stretches the chains.

\begin{figure}
\includegraphics*[width=3in]{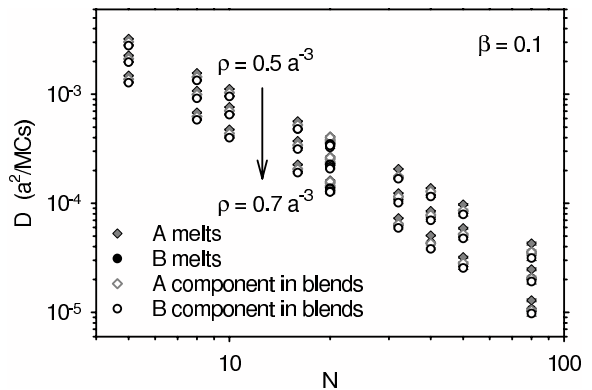}
\caption{\label{figDall} 
Self diffusion coefficients as a function of chain length for all
melts and blends considered in this work. The filled and open symbols
represent simulation results for melts and blends, respectively, while 
diamonds and circles stand for chains of type A and B, respectively. 
The results represent three different monomer densities, 
$\rho=0.5a^{-3}$, $\rho=0.6a^{-3}$, and $\rho=0.7a^{-3}$, and show the
expected decrease of the self-diffusion coefficients with increasing
density. 
In this graph, results for the B melts are hidden by the symbols for
blend results.}
\end{figure}

Self diffusion coefficients give information about the large scale
dynamics of melts and blends.  
They are determined from 
Monte Carlo simulations as described in Section \ref{evaluation}.  
In Fig.~\ref{figDall} we present results for the self-diffusion
coefficients as a function of chain length 
for all melts and blends considered in this work. 
As expected, the self-diffusion coefficients decrease with increasing
monomer density $\rho$ and chain length $N$.  
The chain-length dependence of the self-diffusion coefficients in
Fig.~\ref{figDall} is intermediate between the scaling laws 
$D\sim N^{-1}$ and $D\sim N^{-2}$ expected from Rouse and
reptation theory, respectively.
This is expected since chain-end effects mask the Rouse behavior of
short chains in dense systems 
(see e.g. Refs. \hspace{0em} \onlinecite{pe87,pe94})
while fully entangled behavior is not usually seen for chains of the
relatively short lengths simulated here
\cite{bi97,mu00c,ta00,kr01,pa04b,le05}.  
The graph also shows that the self-diffusion coefficients for A melts
are typically larger than those for the blends. (The symbols for the B
melts are hidden by the symbols for the blends.) 
The differences between the dynamics of the two components and 
the composition dependence of the self-diffusion coefficients will be
discussed in detail for chains of length $N=20$ below. 

\begin{figure}
\includegraphics*[width=3in]{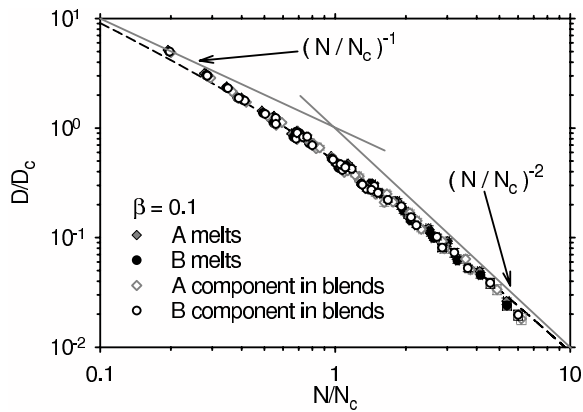}
\caption{\label{figDoDc} 
Scaling representation of the self-diffusion coefficients for all
melts and blends considered in this work. The symbols, which are the
same as in Fig.~\protect{\ref{figDall}} represent values
of the self-diffusion coefficient $D$ divided by
$D_\mathrm{c}=\mu/N_\mathrm{c}$ calculated from
Eqs.~(\protect{\ref{Dc}}), (\protect{\ref{musim}}), and
(\protect{\ref{Ncsims}}) as a function of the chain length $N$ divided
by the characteristic chain length $N_\mathrm{c}$ calculated from 
Eq.~(\protect{\ref{Ncsims}}). 
The dashed line represents the crossover function of Hess
\protect\cite{he86,he88}, see
Eq.~(\protect{\ref{DHess}}), while the gray solid lines represent the
limiting power laws of the Rouse and reptation models
\protect\cite{do86}, see Eq.~(\ref{DbyDc}).   
}
\end{figure}

In Fig.~\ref{figDoDc} we present the self-diffusion coefficient data
of Fig.~\ref{figDall} in scaling form, 
$D/D_\mathrm{c}$ as a function of $N/N_\mathrm{c}$, where 
$D_\mathrm{c}$ is calculated from
Eqs.~(\protect{\ref{Dc}}), (\protect{\ref{musim}}), and
(\protect{\ref{Ncsims}}) and $N_\mathrm{c}$ is calculated from  
Eq.~(\protect{\ref{Ncsims}}). 
In the graph, both melt and blend data for different chain
lengths and densities collapse onto a common line.
This suggests that scaling of dynamic properties with the packing
length may be applicable to blends as well as melts. 
It also gives us some confidence
in our construction of  mobilities from the acceptance rates and
the effective bond lengths ($\mu\sim b^2 A_\text{rate}$). 
A comparison with the gray solid lines indicating the power laws
expected from Rouse and reptation theory \cite{do86} (Eq.~\ref{DbyDc})
confirms that 
most of our data are in the crossover region between unentangled and
fully entangled behavior. 
The dashed line in Fig.~\ref{figDoDc} represents the crossover
equation (\ref{DHess}) of  
Hess \cite{he86,he88}, which was used to determine values for the 
two constants $C_\mu$ and $C_c$ appearing in Eqs.~(\ref{musim})
and (\ref{Ncsims}). 
There is some uncertainty in the results for the constants since more
than one combination of $C_\mu$ and $C_c$ values leads to a respectable
representation of the data in the crossover region. 
However, a difference in a constant prefactor for the mobilities $\mu$
is of little concern since we are interested in the variation with
density and composition rather than the absolute values of the
mobilities and the corresponding friction coefficients
$\zeta=1/\mu\beta$.  

\begin{figure}
\includegraphics*[width=3in]{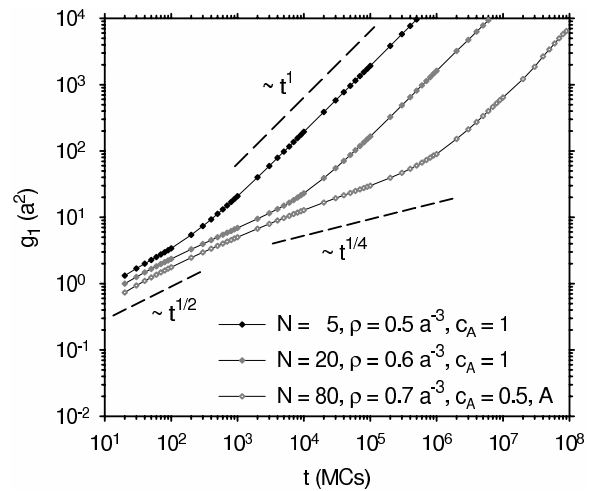}
\caption{\label{figg1}
Average mean squared displacement $g_1$ of the innermost monomer as a
function of time for chains of type A with three different reduced
chain lengths $N/N_\mathrm{c}$. The symbols connected by thin
solid lines represent, from top
to bottom, results for chains of length $N=5$ in a melt of density 
$\rho = 0.5a^{-3}$ ($N/N_\mathrm{c}=0.19$), 
chains of length $N=20$ in a melt of density $\rho = 0.6a^{-3}$
($N/N_\mathrm{c}=0.96$), and 
chains of length $N=80$ in a 50/50 blend of density $\rho = 0.7a^{-3}$
($N/N_\mathrm{c}=6.2$).
The dashed lines indicate the power laws expected for $g_1(t)$ from
reptation theory \protect\cite{do86}.
}
\end{figure}

To help interpret our values for the characteristic chain lengths
$N_\mathrm{c}$ we present in Fig.~\ref{figg1} 
average mean-squared displacements of the central monomer, $g_1(t)$,
for three systems corresponding to the
smallest, the largest and an intermediate value of $N/N_\mathrm{c}$ in
Fig.~\ref{figDoDc}. 
The symbols connected by solid lines represent values calculated
from Eq.~(\ref{g1}) with the block algorithm. 
They represent results for 
an A-melt of chains of length $N=5$ at density $\rho = 0.5a^{-3}$
with $N/N_\mathrm{c}= 0.19$ (top), 
an A-melt of chains of length $N=20$ at density $\rho = 0.6a^{-3}$
with $N/N_\mathrm{c}= 0.96$ (center), and the A component in 
a 50/50 blend of chains of length $N=80$ at density $\rho = 0.7a^{-3}$
with $N/N_\mathrm{c}= 6.2$ (bottom). 
All three examples show the expected diffusive behavior ($g_1\sim t$)
at long times. The behavior at short times is subdiffusive but not
quite Rouse-like ($g_1\sim t^{1/2}$), as is generally observed in
simulations \cite{kr01,pa02b}.  
The largest $g_1(t)$ values correspond to the smallest
reduced chain length, $N/N_\mathrm{c}=0.19$, and illustrate unentangled
behavior; the slope of $g_1(t)$ in this double logarithmic plot 
increases monotonically from the 
subdiffusive region at short times until it reaches a value of unity
in the diffusive region, which starts around $10^3$ MCs for this system.
The smallest $g_1(t)$ values correspond to the largest reduced chain
length, $N/N_\mathrm{c}=6.2$. In this case, 
a second subdiffusive region at intermediate times is clearly
visible. The slope in this region has a value of about 0.37 which is 
not as small as the value of $1/4$ expected for fully
entangled chains from reptation theory \cite{do86}. 
This implies that even the longest chains in our simulations are 
not fully entangled.
The line of intermediate $g_1(t)$ values in Fig.~\ref{figg1}, finally,
corresponds to a melt with chains near the characteristic chain length
($N/N_\mathrm{c}=0.96$). In this case, the slope of $g_1(t)$ decreases
only slightly before it rises to the diffusive value of unity. 
This suggests that the chains are just starting to feel entanglement
effects for reduced chain lengths near $N/N_\mathrm{c}\simeq 1$.

\begin{figure}
\includegraphics*[width=3in]{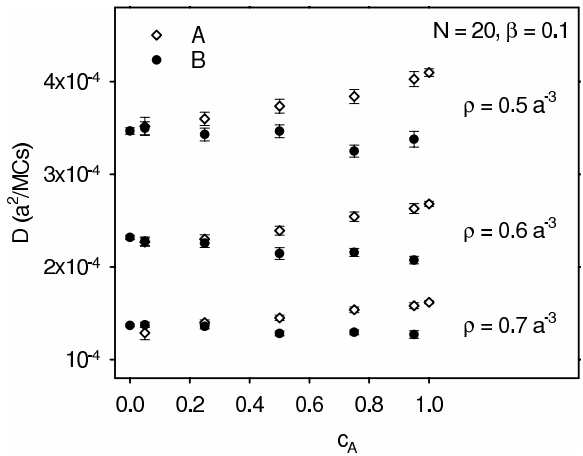}
\caption{\label{figN20D}
Self-diffusion coefficients of melts and blends as a function of blend
composition  
for chains of length $N=20$ and monomer densities 
$\rho = 0.5 A^{-3}$ (top), $\rho = 0.6 A^{-3}$ (center), and $\rho =
0.7 A^{-3}$ (bottom). 
Results for A melts and the A component in blends are represented
by open diamonds, while those for B melts and the B component in
blends are represented by filled circles; 
$c_A$ denotes the mass fraction of component A. 
}
\end{figure}

In order to investigate the composition dependence of the dynamics, we
focus in the following on one chain length, $N=20$, and investigate 
a broad range of compositions ranging from the pure melts
($c_A=0$, $c_A=1$) to the dilute limits ($c_A=0.05$, $c_A=0.95$)
with three intermediate compositions ($c_A = 0.25$, $c_A=0.5$, and
$c_A=0.75$), for the three monomer densities, $\rho = 0.5 a^{-3}$, 
$\rho = 0.6 a^{-3}$, and $\rho = 0.7 a^{-3}$ and reduced inverse
temperature $\beta=0.1$. 
In Fig.~\ref{figN20D} we present the simulation results for the
self-diffusion coefficients of the $N=20$ melts and blends as a
function of composition for the three monomer densities 
$\rho = 0.5 a^{-3}$, $\rho = 0.6 a^{-3}$, and $\rho = 0.7 a^{-3}$. 
These results 
show that, in general, the composition dependence of the
self-diffusion coefficients decreases with increasing density and is
stronger for the A component than for the B component. 
The trends for the self-diffusion coefficients are the same 
at each density; as the mass fraction $c_A$ of the A component
increases, the self-diffusion coefficients of A chains increase while
those of B chains decrease.
Hence, for both types of chains, blending tends to reduce the values of 
the self-diffusion coefficients from the melt values  
(the self-diffusion coefficients of the B components are largest in B
melts and decrease as A component is added, while the $D$ values of 
A chains are largest in the A melt and decrease as B component is
added to the blends).
For most blends considered here, the self-diffusion coefficients of
the A component are larger than those of the B component. 
This is true for all blends of the lowest density considered here,
$\rho=0.5a^{-3}$. However, for the density $\rho=0.6a^{-3}$, the
values of the self-diffusion coefficients of both components are
comparable for blends with small A content. 
For the highest density considered in our simulations,
$\rho=0.7a^{-3}$, finally, the self-diffusion coefficient of the A
component is even smaller than that of the B component for the
smallest blend value of $c_A$. 

\begin{figure}
\includegraphics*[width=3in]{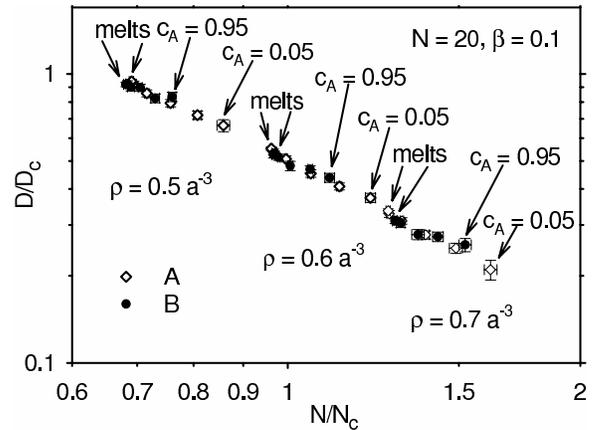}
\caption{\label{figN20DoDc}
Scaling representation of the self-diffusion coefficients for melts
and blends with chain length $N=20$ at the monomer densities
$\rho=0.5 a^{-3}$, $\rho=0.6 a^{-3}$, and $\rho=0.7 a^{-3}$.
The open diamonds and filled circles correspond to results for the A
and B component, respectively. 
For each density, the results for the pure melts correspond to the
lowest reduced chain lengths $N/N_\mathrm{c}$ and the largest values
of the scaled self-diffusion coefficients $D/D_\mathrm{c}$.
Conversely, the largest values of $N/N_\mathrm{c}$ and the smallest
values of $D/D_\mathrm{c}$ for each density correspond to the minority
component in a dilute blend.
}
\end{figure}

In Fig.~\ref{figN20DoDc} we present simulation results of the
self-diffusion coefficients in scaled form, $D/D_\mathrm{c}$ as
a function of $N/N_\mathrm{c}$. The data are a subset of
those presented in Fig.~\ref{figDoDc} and belong to the crossover
region near $N/N_\mathrm{c}=1$. 
In the double-logarithmic graph of Fig.~\ref{figN20DoDc}, 
the reduced self-diffusion coefficient data 
follow closely a straight line  with a slope of about -1.7.  
The reduced diffusion coefficients of the minority
components in dilute blends appear to lie somewhat above this line;  
however, this may not be significant considering the uncertainty in
the data. 
Hence, the data in Fig.~\ref{figN20DoDc} may be approximated
by
\beq\label{DoDcapp}
D/D_\mathrm{c} \simeq \text{const.}\times(N/N_\mathrm{c})^{-\nu} 
\text{ with } \nu\simeq 1.7 ,
\enq
which implies
\beq\label{Dapp}
D \simeq \text{const.}\times\mu N_\mathrm{c}^{\nu-1}  N^{-\nu} ,
\enq
where $D_\mathrm{c}=\mu/N_\mathrm{c}$ of Eq. (\ref{Dc}) has been
used. 
The results presented in Fig.~\ref{figN20DoDc} 
suggest that the scaling with a characteristic chain
length $N_\mathrm{c}$ derived from the packing 
length and a characteristic diffusion coefficient
$D_\mathrm{c}=\mu/N_\mathrm{c}$ constructed with the mobility 
$\mu\sim b^2 A_\text{rate}$ captures both density and
composition dependence of the self diffusion coefficients for these
blends.

\begin{figure}
\includegraphics*[width=3in]{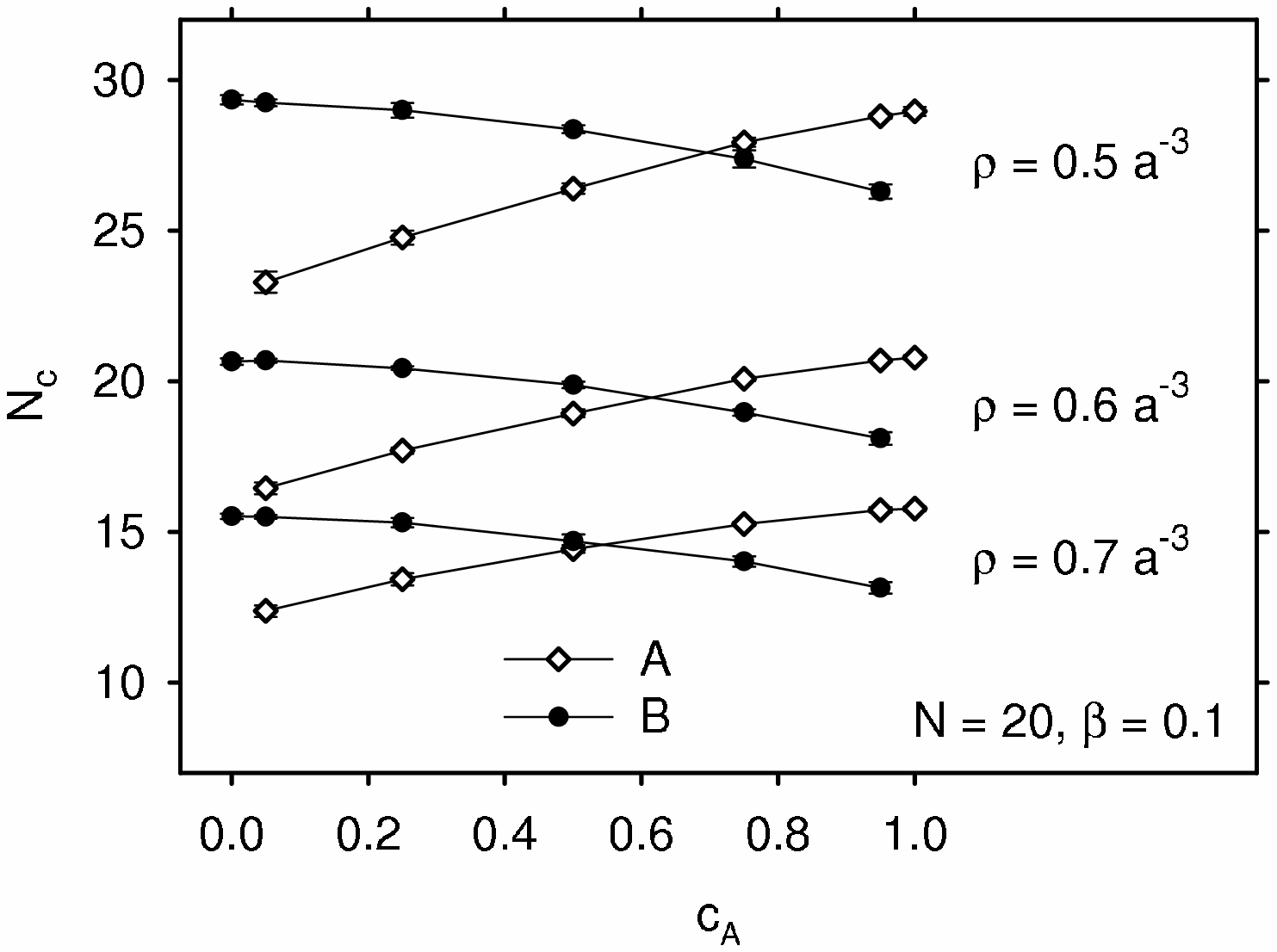}
\caption{\label{figN20Nc}
Scaling chain length $N_\mathrm{c}$ as a function of blend
composition for melts and blends with 
chain length $N=20$ at the indicated densities. 
The symbols  connected by solid lines represent values calculated from
Eq.~(\protect{\ref{Ncsims}}) with Eq.~(\ref{b2}) 
for components A (open diamonds)  
and B (filled circles). 
}
\end{figure}

\begin{figure}
\includegraphics*[width=3in]{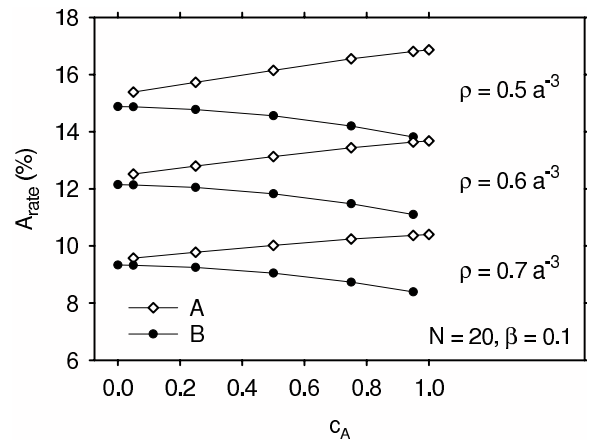}
\caption{\label{figN20Arate}
Average acceptance rates for elementary moves as a function of blend
composition for melts and blends with chain length $N=20$ at the
indicated densities.  
The open diamonds and filled circles connected by solid lines 
represent simulation results for components A and B, respectively.   
}
\end{figure}

\begin{figure}
\includegraphics*[width=3in]{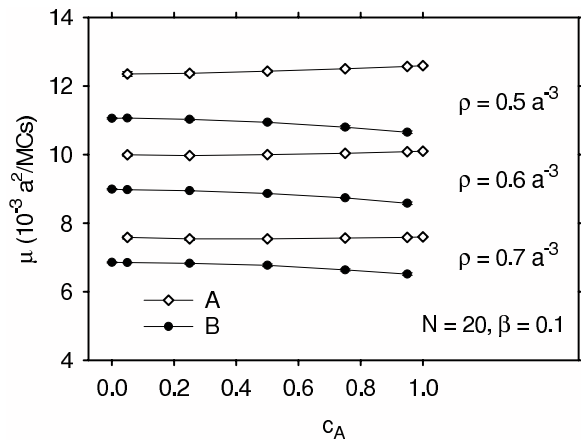}
\caption{\label{figN20mu}
Mobility $\mu$ as a function of blend
composition for melts and blends with chain length $N=20$ at the
indicated densities.  
The symbols connected by solid lines  represent values calculated from
Eq.~(\protect{\ref{musim}}) with Eq.~(\ref{b2}) 
for components A (open diamonds)  
and B (filled circles). 
}
\end{figure}

In order to gain insight into the origin of the variation of the 
self-diffusion coefficients with density and blend composition, we
present in Figs.~\ref{figN20Nc}, \ref{figN20Arate}, and \ref{figN20mu} 
results for the characteristic chain lengths, the acceptance rates,
and the mobilities of the melts and blends with $N=20$. 
The results for the scaling chain lengths $N_\mathrm{c}$ in
Fig.~\ref{figN20Nc} show that $N_\mathrm{c}$ increases with increasing
density. This suggests a decrease of the entanglement length with
increasing density, which is expected since the distance between
monomers from different chains decreases with increasing density
\cite{mi05}.  
The $N_\mathrm{c}$ values in Fig.~\ref{figN20Nc} also
show that blending decreases the 
characteristic chain length for the blends considered here; 
for both types of chain, the values of $N_\mathrm{c}$ decrease with
increasing concentration of the other component at a given density. 
This composition dependence of $N_\mathrm{c}$ is
a result of the variation of the effective bond length $b$ with
composition. The results for $b^2$ in Fig.~\ref{b2} show that both A
and B chains are expanded in a blend compared to the melt. 
Since, according to Eq.~(\ref{Ncsims}), a larger effective bond length
corresponds to a smaller characteristic chain length, the values of
$N_\mathrm{c}$ decrease upon blending from their melt values. 

The average acceptance rates for elementary moves of our Monte Carlo
simulations presented in Fig.~\ref{figN20Arate} measure the average
probability for a monomer to complete an attempted move to a
nearest-neighbor site. In the lattice model for polymer blends
employed in this work, this probability depends, first of all, 
on the local bond structure since only three different bond lengths 
are allowed. It also depends strongly on the local density since the
monomer cannot move to a site that is already occupied by another
monomer. The composition of the blends affects the acceptance rate
through the change in internal energy associated with a move.
The internal energy of the system depends on the numbers 
($N_{ij}$) and the interaction energies ($\epsilon_{ij}$,
$i,j \in \{A,B\}$) of contacts between non-bonded nearest neighbors,
see Eq.~(\ref{E}).   
According to the Metropolis criterion, a move is always accepted when the
new configuration has a lower internal energy than the original
one. If the new internal energy is higher, the move is accepted only
with a probability corresponding to the Boltzmann factor of the energy 
difference between old and new configurations. 
This implies that the acceptance rate for moves is small, when the
new site has a smaller number of occupied nearest neighbor sites or
when the interactions with the new neighbors are less attractive than
those with the old. 
Finally, a small fraction of moves is prohibited because it would lead
to the crossing of bonds.  
The results for the acceptance rates in Fig.~(\ref{figN20Arate}) 
show the expected strong dependence on the monomer density. 
In order to understand the composition dependence, we start by
considering blends, where the A component is dilute, i.e. near
$c_A=0$. In this case, the A chains are surrounded by B chains. 
Since the interaction energies for AB and BB contacts are identical, 
$\epsilon_{AB}=\epsilon_{BB}=-2\epsilon$, 
we expect similar acceptance rates for the A and B component at a
given density. This is indeed what the data in Fig.~\ref{figN20Arate}
show. 
For all compositions and densities, the acceptance rates for the A
component are higher than those of the B component. 
Furthermore, for a given density, as A component is added to a blend, 
the mobility of A monomers increases while that of B monomers
decreases. There are two factors, both related to the energetics, that
contribute to this. First of all, since A-A contacts are less
attractive than A-B contacts 
($\epsilon_{AA}=-1\epsilon$ while
$\epsilon_{AB}=-2\epsilon$), 
A monomers become increasingly more
mobile as A component is added to a blend. Secondly, the local density
near A monomers is somewhat lower than that near B monomers leading to
a larger mobility of A monomers. The reason for the local density
variation is the difference in the average energy penalty for making a 
contact with a void instead of another monomer. 
For B monomers, this penalty is always $2\epsilon$. For A monomers, on
the other hand, it varies with composition between $2\epsilon$ near
$c_A=0$ and $1\epsilon$ near $c_A=1$. 
The effect becomes smaller with increasing monomer density since the
number of voids decreases.

In Fig.~\ref{figN20mu} we present results for the composition
dependence of the mobilities, calculated from
Eq.~(\protect{\ref{musim}}) with Eq.~(\ref{b2}), 
for the melts and blends with chain length $N=20$.  
The mobilities are proportional to both the acceptance rates and
the squared effective bond lengths, $\mu\sim b^2A_\text{rate}$. 
It is not surprising that they show some of the characteristics of the
acceptance rates discussed above. Just like the acceptance rate, the
mobility decreases with increasing density. Furthermore,   
chains of type A always have a larger mobility than chains of type B,
which makes A the faster component in the blend at all compositions. 
However, the composition dependence of the effective bond length
partially compensates that of the acceptance rate and makes the
mobilities less dependent on composition than the acceptance rates. 
As A component is added to a blend, the mobility of A chains
increases, while that of B chains decreases only slightly.  
In the limit where the A component is dilute, the mobilities of the two
components remain distinct, while the acceptance rates of the two
components converge. 
Conversely, in the limit of dilute B in A, the relative differences
between mobilities of the components are smaller than those of the
acceptance rates.

\begin{figure}
\includegraphics*[width=3in]{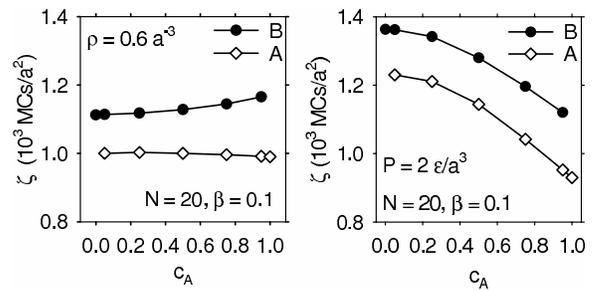}
\caption{\label{figzetas}
Local friction coefficients as a function of
composition for melts and blends with chain length $N=20$. 
The symbols in the left panel represent values $\zeta=1/\beta\mu$ 
corresponding to the mobilities $\mu$ at the constant density 
$\rho=0.6 a^{-3}$ in Fig.~\protect\ref{figN20mu}. 
The symbols in the right panel represent friction coefficient values
interpolated from the data in Fig.~\ref{figzetavsP} to a constant
pressure of $P=2 \epsilon/a^3$. 
The solid lines only connect the symbols. 
}
\end{figure}

\begin{figure}
\includegraphics*[width=3in]{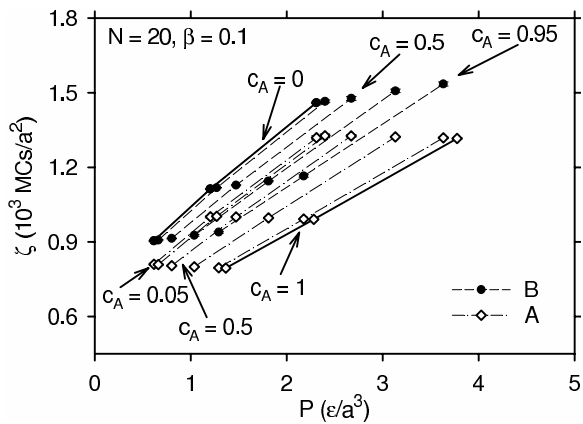}
\caption{\label{figzetavsP}
Local friction coefficients as a function of
pressure for melts and blends with chain length $N=20$.
The symbols represent values of $\zeta=1/\beta\mu$ 
corresponding to the mobilities $\mu$ in Fig.~\protect\ref{figN20mu}. 
The pressure has been estimated from Flory-Huggins theory
\protect\cite{fl53}. 
The dashed and dashed-dotted lines connect values for the B and A
components, respectively, in blends of a given composition;
melt values are connected by solid lines.
The results shown here correspond to mass fractions $c_A=0$,
0.05, 0.25, 0.5, 0.75, 0.95, and 1.0. 
The friction coefficients for the B component in the melt and in the
blend with $c_A=0.05$ are too close to be distinguished in this
representation.  
}
\end{figure}

The local friction coefficients $\zeta$ of the blend components 
are inversely proportional to the mobilities, $\zeta=1/\beta\mu$ 
(Eq.~(\ref{zetasim})).  
In the left panel of Fig.~\ref{figzetas}, we present values for the
local friction coefficients calculated from the mobility results 
for melts and blends of density $\rho=0.6 a^{-3}$ presented in
Fig.~\ref{figN20mu}. 
Since A is the faster component, its friction coefficients are smaller
than those of the B component. 
As the mass fraction of A chains increases, the friction coefficients
of the A component decrease (slightly) while those of the B component
increase.  Hence, when A component is added to a blend at constant
density, A segments speed up (somewhat) while B segments slow
down. 
This is the case for all densities considered here and contrasts
with the experimental observation that both components speed up when
the fraction of faster component in a blend is increased. 
The reason for this apparent discrepancy is that we have varied the
composition at constant
density $\rho$, whereas in experiments the composition is typically
varied at constant pressure. 
In order to arrive at friction coefficient values at constant
pressure, we first consider their pressure dependence and then
interpolate our data to a given pressure.
In this work, we estimate the pressure using a random mixing
approximation for the internal energy and the Flory-Huggins
expression for the entropic contribution \cite{fl53}.  
This is expected to yield at least qualitatively correct results since
our simulations are carried out at high density, where the
Flory-Huggins theory has been found to give a 
reasonable representation of the pressure of athermal lattice chains
\cite{di86,di87}, and since we work at a constant, high temperature
where the random mixing approximation is expected to be adequate. 
In Fig.~\ref{figzetavsP} we present results for the friction
coefficients of melts and blends with $N=20$ as a function of
pressure. As expected, the friction coefficients increase with
increasing pressure for each composition. 
In this representation it becomes apparent that the friction
coefficients of both components decrease when the mass fraction of A
is increased at constant pressure. To illustrate this further, we have
estimated values for friction coefficients at the pressure
$P=2\epsilon/a^3$ by interpolation of the results in
Fig.~\ref{figzetavsP}. 
These values are presented as a function of mass fraction in the right
panel of Fig.~\ref{figzetas}. 
In qualitative agreement with experiment, the results for the
friction coefficients show that the dynamics of both components speed
up when the fraction of the fast component (A) is increased at
constant pressure. 

\section{Discussion}\label{discussion}

In this work, we have presented Monte Carlo simulation results for 
polymer melts and blends with a range of densities, compositions and 
chain lengths.
In order to represent the two components of a binary blend, we have
generalized Shaffer's bond fluctuation model for athermal melts
\cite{sh94,sh95,pa96} by
introducing attractive interactions between occupied non-bonded
nearest neighbor sites on the lattice. 
The interactions parameters employed in this work are 
$\epsilon_{AA}=-1\epsilon$ and $\epsilon_{BB}=-2\epsilon$ for like
interactions between monomers of type A and B, respectively, and
$\epsilon_{AB}=-2\epsilon$ for interactions between unlike monomers. 
All simulations were carried out a fixed
temperature of $T=10\epsilon/\kB$ corresponding to an inverse reduced
temperature of $\beta=0.1$. 
Our results for the effective bond length presented in
Fig.~\ref{figb2} show that the chain dimensions of both types of
chains increase upon blending. This effect increases with increasing
dilution and is due to the large attractive
interaction energy associated with contacts between unlike monomers. 

Self-diffusion coefficients were determined from the mean square
displacement of the center of mass of the chains. 
The simulation results presented in Fig.~\ref{figDall} show that the
chain-length dependence of the self-diffusion coefficients falls
between the power laws expected from Rouse and reptation theory
\cite{do86}, which is expected for the range of chain lengths
considered in this work \cite{bi97,mu00c,ta00,kr01,pa04b,le05}. 
In order to investigate the scaling behavior of the self-diffusion
coefficients, characteristic chain lengths $N_\mathrm{c}$ 
were calculated from the packing lengths of the chains. 
Segmental mobilities $\mu$ were determined from the acceptance rates
and the effective bond lengths and combined with the values for
$N_\mathrm{c}$ to yield characteristic self-diffusion coefficients 
$D_\mathrm{c}=\mu/N_\mathrm{c}$. 
The results for the reduced diffusion
coefficients $D/D_\mathrm{c}$ as a function of the reduced chain
length $N/N_\mathrm{c}$ presented in Figs.~\ref{figDoDc} and
Figs.~\ref{figN20DoDc} show the data for melts and blends with
different chain lengths, densities, and compositions to collapse
onto a common line. This suggests that the 
scaling of dynamic properties  with the packing length, 
which has been observed for polymer melts
\cite{li87,ka87,fe99,ev04,su05,mi05}, 
may be applicable to blends as well. 

The composition dependence of the dynamic properties was studied in
detail for chains of length $N=20$ in melts and blends of three
different densities. The results for the self diffusion coefficients
$D$ presented in Fig.~\ref{figN20D} show the expected variation with
density. At each density and for both types of chains, blending
reduces the self-diffusion coefficient from its melt value.
This may be attributed to the composition dependence of the 
mobility $\mu$ and the characteristic chain length $N_\mathrm{c}$, as
follows. 
The scaling representation of Fig.~\ref{figN20DoDc} for chains of
length $N=20$ shows that the 
reduced self-diffusion coefficient follow approximately
a power law with exponent $\nu=1.7$. 
This implies that $D$ depends on the
mobility $\mu$ and the characteristic chain length $N_\mathrm{c}$ 
approximately as 
$D = \text{const.}\times\mu N_\mathrm{c}^{\nu-1}N^{-\nu}$, see
Eq.~(\ref{Dapp}). 
Since the values of both the mobility and the characteristic chain
length decrease upon blending, so do the values of the self-diffusion
coefficients. 

The variation of the self-diffusion coefficients with 
composition has another interesting aspect.  
For the highest density, $\rho=0.7 a^{-3}$ the difference between the
self-diffusion coefficients of the two components changes 
sign from $D_A-D_B<0$ to $D_A-D_B>0$ at a mass fraction of A of about 
$c_A\simeq 0.2$. 
This is surprising since it implies that, for $\rho=0.7 a^{-3}$,  
A chains move more slowly than B chains at low A concentrations 
but faster than B chains at higher A concentrations. 
Since changes in both mobility and characteristic chain length affect the
value of the diffusion coefficient we consider them in turn. 
Results for the segmental mobility 
presented in Fig.~\ref{figN20mu} show that the local mobility of the A
chains in a blend is always larger than that of the B chains. This
implies that,  on the length scale of an effective bond, 
the A component is always the fast component in the 
blends considered here. Hence, the segmental mobility cannot be
responsible for the sign change of $D_A-D_B$. 
Results for the characteristic chain lengths $N_\mathrm{c}$ 
in Fig.~\ref{figN20Nc} show that, for all densities, the differences
between the characteristic chain lengths change sign from
$N_{cA}-N_{cB}<0$ to $N_{cA}-N_{cB}>0$ at intermediate mass
fractions. 
This, together with the relatively weak composition dependence of
the mobility at $\rho=0.7 a^{-3}$, gives rise to the sign change of
$D_A-D_B$. 
In physical terms; for the highest density considered here and 
at the lowest concentration of A chains, the characteristic chain
length for A chains is so much smaller than that of B chains,  that
entanglement effects lead 
to A chains diffusing more slowly than B chains even though the
segmental mobility of the A chains is larger than that of the B chains. 

Values for the local friction coefficients $\zeta$ have been
determined from the results for the segmental mobilities $\mu$. 
The pressure dependence of the friction coefficients is shown in
Fig.~\ref{figzetavsP}, where the values for the pressure are estimated 
from the Flory-Huggins theory \cite{fl53}. The results show the
expected increase in the friction coefficients with increasing
pressure. The pressure dependence is slightly larger for the
slower (B) component than for the faster (A) component, which agrees
qualitatively with results from our calculations for polyolefin
blends \cite{lu05}. 
In Fig.~\ref{figzetas}, we 
contrast two sets of results for the composition
dependence of the local friction coefficients. 
The results in the left panel, where the
composition of the blends is varied at constant density, as is typical
for simulations,  
are qualitatively different from those in the   
right panel, where the pressure is kept approximately constant, as is 
typical for experiments.  
The results for the constant-pressure composition dependence of the
friction coefficients agree qualitatively with experimental
observations on miscible blends; the friction coefficients of both
components decrease with increasing mass fraction of the fast
component (A).

In this work, we calculate the local friction coefficients from the 
mobility $\mu$, which describes the dynamics on the scale of the
effective bond length. 
For polymer melts near the glass transition \cite{in96} and for
miscible polyolefin blends \cite{ne05}, the Kuhn length has been
identified as the size of the moving segment relevant to local
dynamics. 
For the blend model considered here, the effective bond length and the
Kuhn length are comparable in size, which makes it difficult to distinguish
between these choices for the size of the moving segments. 
However, we hope that a current investigation of the dynamics
of athermal melts for a larger set of densities will give us some
insight into this question \cite{lu06}.

It is encouraging that these first simulation results for our
lattice model for polymer blends reproduce trends observed in
experimental work on polymer blends. The differences between the
friction coefficients of the two melts and between the melts and the
blends, however, are fairly small. The reason is that 
chains of type A and B differ only in the values of the interaction
parameters and that the simulations are carried out at high
temperature. 
In real polymer blends, the blend components typically differ in chain
stiffness, which is known to have a large effect on the
dynamics \cite{fe99,mi05,mu00c,ka03b}.  
In order to simulate more realistic systems and to 
increase the difference between the chains of the blend components, 
we are currently extending our model to include a differences in
chain stiffness.  

\section*{Acknowledgments}
Financial support through the National Science Foundation (NSF
DMR-0103704), the Research Corporation (CC5228), and the Petroleum
Research Fund (PRF \#364559 GB7) is gratefully acknowledged.

\bibliography{../mybib/dart}

\begin{thebibliography}{64}
\expandafter\ifx\csname natexlab\endcsname\relax\def\natexlab#1{#1}\fi
\expandafter\ifx\csname bibnamefont\endcsname\relax
  \def\bibnamefont#1{#1}\fi
\expandafter\ifx\csname bibfnamefont\endcsname\relax
  \def\bibfnamefont#1{#1}\fi
\expandafter\ifx\csname citenamefont\endcsname\relax
  \def\citenamefont#1{#1}\fi
\expandafter\ifx\csname url\endcsname\relax
  \def\url#1{\texttt{#1}}\fi
\expandafter\ifx\csname urlprefix\endcsname\relax\def\urlprefix{URL }\fi
\providecommand{\bibinfo}[2]{#2}
\providecommand{\eprint}[2][]{\url{#2}}

\bibitem[{\citenamefont{Doi and Edwards}(1986)}]{do86}
\bibinfo{author}{\bibfnamefont{M.}~\bibnamefont{Doi}} \bibnamefont{and}
  \bibinfo{author}{\bibfnamefont{S.~F.} \bibnamefont{Edwards}},
  \emph{\bibinfo{title}{The Theory of Polymer Dynamics}}
  (\bibinfo{publisher}{Clarendon}, \bibinfo{address}{Oxford},
  \bibinfo{year}{1986}).

\bibitem[{\citenamefont{Grosberg and Khokhlov}(1994)}]{gr94c}
\bibinfo{author}{\bibfnamefont{A.~Y.} \bibnamefont{Grosberg}} \bibnamefont{and}
  \bibinfo{author}{\bibfnamefont{A.~R.} \bibnamefont{Khokhlov}},
  \emph{\bibinfo{title}{Statistical Physics of Macromolecules}}, AIP Series in
  Polymers and Complex Materials (\bibinfo{publisher}{American Institute of
  Physics}, \bibinfo{address}{Woodbury, NY}, \bibinfo{year}{1994}).

\bibitem[{\citenamefont{de~Gennes}(1979)}]{de79}
\bibinfo{author}{\bibfnamefont{P.-G.} \bibnamefont{de~Gennes}},
  \emph{\bibinfo{title}{Scaling Concepts in Polymer Physics}}
  (\bibinfo{publisher}{Cornell University}, \bibinfo{address}{Ithaca, NY},
  \bibinfo{year}{1979}).

\bibitem[{\citenamefont{Ferry}(1980)}]{fe80}
\bibinfo{author}{\bibfnamefont{J.~D.} \bibnamefont{Ferry}},
  \emph{\bibinfo{title}{Viscoelastic Properties of Polymers}}
  (\bibinfo{publisher}{Wiley}, \bibinfo{address}{New York},
  \bibinfo{year}{1980}), \bibinfo{edition}{3rd} ed.

\bibitem[{\citenamefont{Roland and Ngai}(1991)}]{ro91b}
\bibinfo{author}{\bibfnamefont{C.~M.} \bibnamefont{Roland}} \bibnamefont{and}
  \bibinfo{author}{\bibfnamefont{K.~L.} \bibnamefont{Ngai}},
  \bibinfo{journal}{Macromolecules} \textbf{\bibinfo{volume}{24}},
  \bibinfo{pages}{2261} (\bibinfo{year}{1991}).

\bibitem[{\citenamefont{Roland and Ngai}(1992)}]{ro92}
\bibinfo{author}{\bibfnamefont{C.~M.} \bibnamefont{Roland}} \bibnamefont{and}
  \bibinfo{author}{\bibfnamefont{K.~L.} \bibnamefont{Ngai}},
  \bibinfo{journal}{Macromolecules} \textbf{\bibinfo{volume}{25}},
  \bibinfo{pages}{363} (\bibinfo{year}{1992}), \bibinfo{note}{{M}acromolecules
  {\bf 33}, 3184 (2000)}.

\bibitem[{\citenamefont{Chung et~al.}(1994{\natexlab{a}})\citenamefont{Chung,
  Kornfield, and Smith}}]{ch94}
\bibinfo{author}{\bibfnamefont{G.~C.} \bibnamefont{Chung}},
  \bibinfo{author}{\bibfnamefont{J.~A.} \bibnamefont{Kornfield}},
  \bibnamefont{and} \bibinfo{author}{\bibfnamefont{S.~D.} \bibnamefont{Smith}},
  \bibinfo{journal}{Macromolecules} \textbf{\bibinfo{volume}{27}},
  \bibinfo{pages}{964} (\bibinfo{year}{1994}{\natexlab{a}}).

\bibitem[{\citenamefont{He et~al.}(2003)\citenamefont{He, Lutz, and
  Ediger}}]{he03}
\bibinfo{author}{\bibfnamefont{Y.}~\bibnamefont{He}},
  \bibinfo{author}{\bibfnamefont{T.~R.} \bibnamefont{Lutz}}, \bibnamefont{and}
  \bibinfo{author}{\bibfnamefont{M.~D.} \bibnamefont{Ediger}},
  \bibinfo{journal}{J. Chem. Phys.} \textbf{\bibinfo{volume}{119}},
  \bibinfo{pages}{9956} (\bibinfo{year}{2003}).

\bibitem[{\citenamefont{Chung et~al.}(1994{\natexlab{b}})\citenamefont{Chung,
  Kornfield, and Smith}}]{ch94b}
\bibinfo{author}{\bibfnamefont{G.~C.} \bibnamefont{Chung}},
  \bibinfo{author}{\bibfnamefont{J.~A.} \bibnamefont{Kornfield}},
  \bibnamefont{and} \bibinfo{author}{\bibfnamefont{S.~D.} \bibnamefont{Smith}},
  \bibinfo{journal}{Macromolecules} \textbf{\bibinfo{volume}{27}},
  \bibinfo{pages}{5729} (\bibinfo{year}{1994}{\natexlab{b}}).

\bibitem[{\citenamefont{Al{\'{e}}gria et~al.}(1994)\citenamefont{Al{\'{e}}gria,
  Colmenero, Ngai, and Roland}}]{al94}
\bibinfo{author}{\bibfnamefont{A.}~\bibnamefont{Al{\'{e}}gria}},
  \bibinfo{author}{\bibfnamefont{J.}~\bibnamefont{Colmenero}},
  \bibinfo{author}{\bibfnamefont{K.~L.} \bibnamefont{Ngai}}, \bibnamefont{and}
  \bibinfo{author}{\bibfnamefont{C.~M.} \bibnamefont{Roland}},
  \bibinfo{journal}{Macromolecules} \textbf{\bibinfo{volume}{27}},
  \bibinfo{pages}{4486} (\bibinfo{year}{1994}).

\bibitem[{\citenamefont{Kim et~al.}(1995)\citenamefont{Kim, Kramer, and
  Osby}}]{ki95}
\bibinfo{author}{\bibfnamefont{E.}~\bibnamefont{Kim}},
  \bibinfo{author}{\bibfnamefont{E.~J.} \bibnamefont{Kramer}},
  \bibnamefont{and} \bibinfo{author}{\bibfnamefont{J.~O.} \bibnamefont{Osby}},
  \bibinfo{journal}{Macromolecules} \textbf{\bibinfo{volume}{28}},
  \bibinfo{pages}{1979} (\bibinfo{year}{1995}).

\bibitem[{\citenamefont{Pathak}(2001)}]{pa01}
\bibinfo{author}{\bibfnamefont{J.~A.} \bibnamefont{Pathak}}, Ph.D. thesis,
  \bibinfo{school}{The Pennsylvania State University} (\bibinfo{year}{2001}).

\bibitem[{\citenamefont{Min et~al.}(2001)\citenamefont{Min, Qiu, Ediger,
  Pitsikalis, and Hadjichristidis}}]{mi01}
\bibinfo{author}{\bibfnamefont{B.}~\bibnamefont{Min}},
  \bibinfo{author}{\bibfnamefont{X.~H.} \bibnamefont{Qiu}},
  \bibinfo{author}{\bibfnamefont{M.~D.} \bibnamefont{Ediger}},
  \bibinfo{author}{\bibfnamefont{M.}~\bibnamefont{Pitsikalis}},
  \bibnamefont{and}
  \bibinfo{author}{\bibfnamefont{N.}~\bibnamefont{Hadjichristidis}},
  \bibinfo{journal}{Macromolecules} \textbf{\bibinfo{volume}{34}},
  \bibinfo{pages}{4466} (\bibinfo{year}{2001}).

\bibitem[{\citenamefont{Haley et~al.}(2003)\citenamefont{Haley, Lodge, He,
  Ediger, von Meerwall, and Mijovic}}]{ha03}
\bibinfo{author}{\bibfnamefont{J.~C.} \bibnamefont{Haley}},
  \bibinfo{author}{\bibfnamefont{T.~P.} \bibnamefont{Lodge}},
  \bibinfo{author}{\bibfnamefont{Y.}~\bibnamefont{He}},
  \bibinfo{author}{\bibfnamefont{M.~D.} \bibnamefont{Ediger}},
  \bibinfo{author}{\bibfnamefont{E.~D.} \bibnamefont{von Meerwall}},
  \bibnamefont{and} \bibinfo{author}{\bibfnamefont{J.}~\bibnamefont{Mijovic}},
  \bibinfo{journal}{Macromolecules} \textbf{\bibinfo{volume}{36}},
  \bibinfo{pages}{6142} (\bibinfo{year}{2003}).

\bibitem[{\citenamefont{Lutz et~al.}(2003)\citenamefont{Lutz, He, Ediger, Cao,
  Lin, and Jones}}]{lu03d}
\bibinfo{author}{\bibfnamefont{T.~R.} \bibnamefont{Lutz}},
  \bibinfo{author}{\bibfnamefont{Y.}~\bibnamefont{He}},
  \bibinfo{author}{\bibfnamefont{M.~D.} \bibnamefont{Ediger}},
  \bibinfo{author}{\bibfnamefont{H.}~\bibnamefont{Cao}},
  \bibinfo{author}{\bibfnamefont{G.}~\bibnamefont{Lin}}, \bibnamefont{and}
  \bibinfo{author}{\bibfnamefont{A.~A.} \bibnamefont{Jones}},
  \bibinfo{journal}{Macromolecules} \textbf{\bibinfo{volume}{36}},
  \bibinfo{pages}{1724} (\bibinfo{year}{2003}).

\bibitem[{\citenamefont{Haley and Lodge}(2004)}]{ha04}
\bibinfo{author}{\bibfnamefont{J.~C.} \bibnamefont{Haley}} \bibnamefont{and}
  \bibinfo{author}{\bibfnamefont{T.~P.} \bibnamefont{Lodge}},
  \bibinfo{journal}{Colloid Polym. Sci.} \textbf{\bibinfo{volume}{282}},
  \bibinfo{pages}{793} (\bibinfo{year}{2004}).

\bibitem[{\citenamefont{Lutz et~al.}(2004)\citenamefont{Lutz, He, Ediger,
  Pitsikalis, and Hadjichristidis}}]{lu04}
\bibinfo{author}{\bibfnamefont{T.~R.} \bibnamefont{Lutz}},
  \bibinfo{author}{\bibfnamefont{Y.}~\bibnamefont{He}},
  \bibinfo{author}{\bibfnamefont{M.~D.} \bibnamefont{Ediger}},
  \bibinfo{author}{\bibfnamefont{M.}~\bibnamefont{Pitsikalis}},
  \bibnamefont{and}
  \bibinfo{author}{\bibfnamefont{N.}~\bibnamefont{Hadjichristidis}},
  \bibinfo{journal}{Macromolecules} \textbf{\bibinfo{volume}{37}},
  \bibinfo{pages}{6440} (\bibinfo{year}{2004}).

\bibitem[{\citenamefont{Alegr{\'{i}}a et~al.}(2002)\citenamefont{Alegr{\'{i}}a,
  G{\'{o}}mez, and Colmenero}}]{al02}
\bibinfo{author}{\bibfnamefont{A.}~\bibnamefont{Alegr{\'{i}}a}},
  \bibinfo{author}{\bibfnamefont{D.}~\bibnamefont{G{\'{o}}mez}},
  \bibnamefont{and}
  \bibinfo{author}{\bibfnamefont{J.}~\bibnamefont{Colmenero}},
  \bibinfo{journal}{Macromolecules} \textbf{\bibinfo{volume}{35}},
  \bibinfo{pages}{2030} (\bibinfo{year}{2002}).

\bibitem[{\citenamefont{Doxastakis et~al.}(2000)\citenamefont{Doxastakis,
  Kitsiou, Fytas, Theodorou, Hadjichristis, Meier, and Frick}}]{do00}
\bibinfo{author}{\bibfnamefont{M.}~\bibnamefont{Doxastakis}},
  \bibinfo{author}{\bibfnamefont{M.}~\bibnamefont{Kitsiou}},
  \bibinfo{author}{\bibfnamefont{G.}~\bibnamefont{Fytas}},
  \bibinfo{author}{\bibfnamefont{D.~N.} \bibnamefont{Theodorou}},
  \bibinfo{author}{\bibfnamefont{N.}~\bibnamefont{Hadjichristis}},
  \bibinfo{author}{\bibfnamefont{G.}~\bibnamefont{Meier}}, \bibnamefont{and}
  \bibinfo{author}{\bibfnamefont{B.}~\bibnamefont{Frick}}, \bibinfo{journal}{J.
  Chem. Phys.} \textbf{\bibinfo{volume}{112}}, \bibinfo{pages}{8687}
  (\bibinfo{year}{2000}).

\bibitem[{\citenamefont{Genix et~al.}(2005)\citenamefont{Genix, Arbe, Alvarez,
  Colmenero, Willner, and Richter}}]{ge05}
\bibinfo{author}{\bibfnamefont{A.-C.} \bibnamefont{Genix}},
  \bibinfo{author}{\bibfnamefont{A.}~\bibnamefont{Arbe}},
  \bibinfo{author}{\bibfnamefont{F.}~\bibnamefont{Alvarez}},
  \bibinfo{author}{\bibfnamefont{J.}~\bibnamefont{Colmenero}},
  \bibinfo{author}{\bibfnamefont{L.}~\bibnamefont{Willner}}, \bibnamefont{and}
  \bibinfo{author}{\bibfnamefont{D.}~\bibnamefont{Richter}},
  \bibinfo{journal}{Phys. Rev. E} \textbf{\bibinfo{volume}{72}},
  \bibinfo{pages}{031808} (\bibinfo{year}{2005}).

\bibitem[{\citenamefont{Kopf et~al.}(1997)\citenamefont{Kopf, D{\"{u}}nweg, and
  Paul}}]{ko97c}
\bibinfo{author}{\bibfnamefont{A.}~\bibnamefont{Kopf}},
  \bibinfo{author}{\bibfnamefont{B.}~\bibnamefont{D{\"{u}}nweg}},
  \bibnamefont{and} \bibinfo{author}{\bibfnamefont{W.}~\bibnamefont{Paul}},
  \bibinfo{journal}{J. Chem. Phys.} \textbf{\bibinfo{volume}{107}},
  \bibinfo{pages}{6945} (\bibinfo{year}{1997}).

\bibitem[{\citenamefont{Kamath et~al.}(2003)\citenamefont{Kamath, Colby, and
  Kumar}}]{ka03b}
\bibinfo{author}{\bibfnamefont{S.}~\bibnamefont{Kamath}},
  \bibinfo{author}{\bibfnamefont{R.~H.} \bibnamefont{Colby}}, \bibnamefont{and}
  \bibinfo{author}{\bibfnamefont{S.~K.} \bibnamefont{Kumar}},
  \bibinfo{journal}{Macromolecules} \textbf{\bibinfo{volume}{36}},
  \bibinfo{pages}{8567} (\bibinfo{year}{2003}).

\bibitem[{\citenamefont{Budzien et~al.}(2002)\citenamefont{Budzien, Raphael,
  Ediger, and de~Pablo}}]{bu02}
\bibinfo{author}{\bibfnamefont{J.}~\bibnamefont{Budzien}},
  \bibinfo{author}{\bibfnamefont{C.}~\bibnamefont{Raphael}},
  \bibinfo{author}{\bibfnamefont{M.~D.} \bibnamefont{Ediger}},
  \bibnamefont{and} \bibinfo{author}{\bibfnamefont{J.~J.}
  \bibnamefont{de~Pablo}}, \bibinfo{journal}{J. Chem. Phys.}
  \textbf{\bibinfo{volume}{116}}, \bibinfo{pages}{8209} (\bibinfo{year}{2002}).

\bibitem[{\citenamefont{Faller}(2004)}]{fa04}
\bibinfo{author}{\bibfnamefont{R.}~\bibnamefont{Faller}},
  \bibinfo{journal}{Macromolecules} \textbf{\bibinfo{volume}{37}},
  \bibinfo{pages}{1095} (\bibinfo{year}{2004}).

\bibitem[{\citenamefont{Neelakantan et~al.}(2005)\citenamefont{Neelakantan,
  May, and Maranas}}]{ne05}
\bibinfo{author}{\bibfnamefont{A.}~\bibnamefont{Neelakantan}},
  \bibinfo{author}{\bibfnamefont{A.}~\bibnamefont{May}}, \bibnamefont{and}
  \bibinfo{author}{\bibfnamefont{J.~K.} \bibnamefont{Maranas}},
  \bibinfo{journal}{Macromolecules} \textbf{\bibinfo{volume}{38}},
  \bibinfo{pages}{6598} (\bibinfo{year}{2005}).

\bibitem[{\citenamefont{Zetsche and Fischer}(1994)}]{ze94}
\bibinfo{author}{\bibfnamefont{A.}~\bibnamefont{Zetsche}} \bibnamefont{and}
  \bibinfo{author}{\bibfnamefont{E.~W.} \bibnamefont{Fischer}},
  \bibinfo{journal}{Acta Polymerica} \textbf{\bibinfo{volume}{45}},
  \bibinfo{pages}{168} (\bibinfo{year}{1994}).

\bibitem[{\citenamefont{Katana et~al.}(1995)\citenamefont{Katana, Fischer,
  Hack, Abetz, and Kremer}}]{ka95b}
\bibinfo{author}{\bibfnamefont{G.}~\bibnamefont{Katana}},
  \bibinfo{author}{\bibfnamefont{E.~W.} \bibnamefont{Fischer}},
  \bibinfo{author}{\bibfnamefont{T.}~\bibnamefont{Hack}},
  \bibinfo{author}{\bibfnamefont{V.}~\bibnamefont{Abetz}}, \bibnamefont{and}
  \bibinfo{author}{\bibfnamefont{F.}~\bibnamefont{Kremer}},
  \bibinfo{journal}{Macromolecules} \textbf{\bibinfo{volume}{28}},
  \bibinfo{pages}{2714} (\bibinfo{year}{1995}).

\bibitem[{\citenamefont{Kumar et~al.}(1996)\citenamefont{Kumar, Colby,
  Anastasiadis, and Fytas}}]{ku96}
\bibinfo{author}{\bibfnamefont{S.~K.} \bibnamefont{Kumar}},
  \bibinfo{author}{\bibfnamefont{R.~H.} \bibnamefont{Colby}},
  \bibinfo{author}{\bibfnamefont{S.~H.} \bibnamefont{Anastasiadis}},
  \bibnamefont{and} \bibinfo{author}{\bibfnamefont{G.}~\bibnamefont{Fytas}},
  \bibinfo{journal}{J. Chem. Phys.} \textbf{\bibinfo{volume}{105}},
  \bibinfo{pages}{3777} (\bibinfo{year}{1996}).

\bibitem[{\citenamefont{Kamath et~al.}(1999)\citenamefont{Kamath, Colby, Kumar,
  Karatasos, Floudas, Fytas, and Roovers}}]{ka99}
\bibinfo{author}{\bibfnamefont{S.}~\bibnamefont{Kamath}},
  \bibinfo{author}{\bibfnamefont{R.~H.} \bibnamefont{Colby}},
  \bibinfo{author}{\bibfnamefont{S.~K.} \bibnamefont{Kumar}},
  \bibinfo{author}{\bibfnamefont{K.}~\bibnamefont{Karatasos}},
  \bibinfo{author}{\bibfnamefont{G.}~\bibnamefont{Floudas}},
  \bibinfo{author}{\bibfnamefont{G.}~\bibnamefont{Fytas}}, \bibnamefont{and}
  \bibinfo{author}{\bibfnamefont{J.~E.~L.} \bibnamefont{Roovers}},
  \bibinfo{journal}{J. Chem. Phys.} \textbf{\bibinfo{volume}{111}},
  \bibinfo{pages}{6121} (\bibinfo{year}{1999}).

\bibitem[{\citenamefont{Lodge and McLeish}(2000)}]{lo00}
\bibinfo{author}{\bibfnamefont{T.~P.} \bibnamefont{Lodge}} \bibnamefont{and}
  \bibinfo{author}{\bibfnamefont{T.~C.~B.} \bibnamefont{McLeish}},
  \bibinfo{journal}{Macromolecules} \textbf{\bibinfo{volume}{33}},
  \bibinfo{pages}{5278} (\bibinfo{year}{2000}).

\bibitem[{\citenamefont{Leroy et~al.}(2003)\citenamefont{Leroy, Alegr{\'{i}}a,
  and Colmenero}}]{le03}
\bibinfo{author}{\bibfnamefont{E.}~\bibnamefont{Leroy}},
  \bibinfo{author}{\bibfnamefont{A.}~\bibnamefont{Alegr{\'{i}}a}},
  \bibnamefont{and}
  \bibinfo{author}{\bibfnamefont{J.}~\bibnamefont{Colmenero}},
  \bibinfo{journal}{Macromolecules} \textbf{\bibinfo{volume}{36}},
  \bibinfo{pages}{7280} (\bibinfo{year}{2003}).

\bibitem[{\citenamefont{Kant et~al.}(2003)\citenamefont{Kant, Kumar, and
  Colby}}]{ka03c}
\bibinfo{author}{\bibfnamefont{R.}~\bibnamefont{Kant}},
  \bibinfo{author}{\bibfnamefont{S.~K.} \bibnamefont{Kumar}}, \bibnamefont{and}
  \bibinfo{author}{\bibfnamefont{R.~H.} \bibnamefont{Colby}},
  \bibinfo{journal}{Macromolecules} \textbf{\bibinfo{volume}{36}},
  \bibinfo{pages}{10087} (\bibinfo{year}{2003}).

\bibitem[{\citenamefont{Ngai and Roland}(2004)}]{ng04}
\bibinfo{author}{\bibfnamefont{K.~L.} \bibnamefont{Ngai}} \bibnamefont{and}
  \bibinfo{author}{\bibfnamefont{C.~M.} \bibnamefont{Roland}},
  \bibinfo{journal}{Rubber Chem. Technol.} \textbf{\bibinfo{volume}{77}},
  \bibinfo{pages}{579} (\bibinfo{year}{2004}).

\bibitem[{\citenamefont{Luettmer-Strathmann}(2005)}]{lu05}
\bibinfo{author}{\bibfnamefont{J.}~\bibnamefont{Luettmer-Strathmann}},
  \bibinfo{journal}{J. Chem. Phys.} \textbf{\bibinfo{volume}{123}},
  \bibinfo{pages}{014910} (\bibinfo{year}{2005}).

\bibitem[{\citenamefont{Graessley and Edwards}(1981)}]{gr81}
\bibinfo{author}{\bibfnamefont{W.~W.} \bibnamefont{Graessley}}
  \bibnamefont{and} \bibinfo{author}{\bibfnamefont{S.~F.}
  \bibnamefont{Edwards}}, \bibinfo{journal}{Polymer}
  \textbf{\bibinfo{volume}{22}}, \bibinfo{pages}{1329} (\bibinfo{year}{1981}).

\bibitem[{\citenamefont{Lin}(1987)}]{li87}
\bibinfo{author}{\bibfnamefont{Y.-H.} \bibnamefont{Lin}},
  \bibinfo{journal}{Macromolecules} \textbf{\bibinfo{volume}{20}},
  \bibinfo{pages}{3080} (\bibinfo{year}{1987}).

\bibitem[{\citenamefont{Kavassalis and Noolandi}(1987)}]{ka87}
\bibinfo{author}{\bibfnamefont{T.~A.} \bibnamefont{Kavassalis}}
  \bibnamefont{and} \bibinfo{author}{\bibfnamefont{J.}~\bibnamefont{Noolandi}},
  \bibinfo{journal}{Phys. Rev. Lett.} \textbf{\bibinfo{volume}{59}},
  \bibinfo{pages}{2674} (\bibinfo{year}{1987}).

\bibitem[{\citenamefont{Fetters et~al.}(1999)\citenamefont{Fetters, Lohse,
  Milner, and Graessley}}]{fe99}
\bibinfo{author}{\bibfnamefont{L.~J.} \bibnamefont{Fetters}},
  \bibinfo{author}{\bibfnamefont{D.~J.} \bibnamefont{Lohse}},
  \bibinfo{author}{\bibfnamefont{S.~T.} \bibnamefont{Milner}},
  \bibnamefont{and} \bibinfo{author}{\bibfnamefont{W.~W.}
  \bibnamefont{Graessley}}, \bibinfo{journal}{Macromolecules}
  \textbf{\bibinfo{volume}{32}}, \bibinfo{pages}{6847} (\bibinfo{year}{1999}).

\bibitem[{\citenamefont{Everaers et~al.}(2004)\citenamefont{Everaers,
  Sukumaran, Grest, Svaneborg, Sivasubramanian, and Kremer}}]{ev04}
\bibinfo{author}{\bibfnamefont{R.}~\bibnamefont{Everaers}},
  \bibinfo{author}{\bibfnamefont{S.~K.} \bibnamefont{Sukumaran}},
  \bibinfo{author}{\bibfnamefont{G.~S.} \bibnamefont{Grest}},
  \bibinfo{author}{\bibfnamefont{C.}~\bibnamefont{Svaneborg}},
  \bibinfo{author}{\bibfnamefont{A.}~\bibnamefont{Sivasubramanian}},
  \bibnamefont{and} \bibinfo{author}{\bibfnamefont{K.}~\bibnamefont{Kremer}},
  \bibinfo{journal}{Science} \textbf{\bibinfo{volume}{303}},
  \bibinfo{pages}{823} (\bibinfo{year}{2004}).

\bibitem[{\citenamefont{Sukumaran et~al.}(2005)\citenamefont{Sukumaran, Grest,
  Kremer, and Everaers}}]{su05}
\bibinfo{author}{\bibfnamefont{S.~K.} \bibnamefont{Sukumaran}},
  \bibinfo{author}{\bibfnamefont{G.~S.} \bibnamefont{Grest}},
  \bibinfo{author}{\bibfnamefont{K.}~\bibnamefont{Kremer}}, \bibnamefont{and}
  \bibinfo{author}{\bibfnamefont{R.}~\bibnamefont{Everaers}},
  \bibinfo{journal}{J. Polym. Sci. Part B: Polym. Phys.}
  \textbf{\bibinfo{volume}{43}}, \bibinfo{pages}{917} (\bibinfo{year}{2005}).

\bibitem[{\citenamefont{Milner}(2005)}]{mi05}
\bibinfo{author}{\bibfnamefont{S.~T.} \bibnamefont{Milner}},
  \bibinfo{journal}{Macromolecules} \textbf{\bibinfo{volume}{38}},
  \bibinfo{pages}{4929} (\bibinfo{year}{2005}).

\bibitem[{\citenamefont{Binder and Paul}(1997)}]{bi97}
\bibinfo{author}{\bibfnamefont{K.}~\bibnamefont{Binder}} \bibnamefont{and}
  \bibinfo{author}{\bibfnamefont{W.}~\bibnamefont{Paul}}, \bibinfo{journal}{J.
  Polym. Sci. B: Polym. Phys.} \textbf{\bibinfo{volume}{35}},
  \bibinfo{pages}{1} (\bibinfo{year}{1997}).

\bibitem[{\citenamefont{M{\"{u}}ller et~al.}(2000)\citenamefont{M{\"{u}}ller,
  Wittmer, and Barrat}}]{mu00c}
\bibinfo{author}{\bibfnamefont{M.}~\bibnamefont{M{\"{u}}ller}},
  \bibinfo{author}{\bibfnamefont{J.~P.} \bibnamefont{Wittmer}},
  \bibnamefont{and} \bibinfo{author}{\bibfnamefont{J.-L.}
  \bibnamefont{Barrat}}, \bibinfo{journal}{Europhys. Lett.}
  \textbf{\bibinfo{volume}{52}}, \bibinfo{pages}{406} (\bibinfo{year}{2000}).

\bibitem[{\citenamefont{Tanaka et~al.}(2000)\citenamefont{Tanaka, Iwata, and
  Kuzuu}}]{ta00}
\bibinfo{author}{\bibfnamefont{M.}~\bibnamefont{Tanaka}},
  \bibinfo{author}{\bibfnamefont{K.}~\bibnamefont{Iwata}}, \bibnamefont{and}
  \bibinfo{author}{\bibfnamefont{N.}~\bibnamefont{Kuzuu}},
  \bibinfo{journal}{Comput. Theor. Polym. Sci.} \textbf{\bibinfo{volume}{10}},
  \bibinfo{pages}{299} (\bibinfo{year}{2000}).

\bibitem[{\citenamefont{Kreer et~al.}(2001)\citenamefont{Kreer, Baschnagel,
  M{\"{u}}ller, and Binder}}]{kr01}
\bibinfo{author}{\bibfnamefont{T.}~\bibnamefont{Kreer}},
  \bibinfo{author}{\bibfnamefont{J.}~\bibnamefont{Baschnagel}},
  \bibinfo{author}{\bibfnamefont{M.}~\bibnamefont{M{\"{u}}ller}},
  \bibnamefont{and} \bibinfo{author}{\bibfnamefont{K.}~\bibnamefont{Binder}},
  \bibinfo{journal}{Macromolecules} \textbf{\bibinfo{volume}{34}},
  \bibinfo{pages}{1105} (\bibinfo{year}{2001}).

\bibitem[{\citenamefont{Paul and Smith}(2004)}]{pa04b}
\bibinfo{author}{\bibfnamefont{W.}~\bibnamefont{Paul}} \bibnamefont{and}
  \bibinfo{author}{\bibfnamefont{G.~D.} \bibnamefont{Smith}},
  \bibinfo{journal}{Rep. Prog. Phys.} \textbf{\bibinfo{volume}{67}},
  \bibinfo{pages}{1117} (\bibinfo{year}{2004}).

\bibitem[{\citenamefont{Le{\'{o}}n et~al.}(2005)\citenamefont{L{'{e}}on, van~der
  Vegt, Delle~Site, and Kremer}}]{le05}
\bibinfo{author}{\bibfnamefont{S.}~\bibnamefont{Le{\'{o}}n}},
  \bibinfo{author}{\bibfnamefont{N.}~\bibnamefont{van~der Vegt}},
  \bibinfo{author}{\bibfnamefont{L.}~\bibnamefont{Delle~Site}},
  \bibnamefont{and} \bibinfo{author}{\bibfnamefont{K.}~\bibnamefont{Kremer}},
  \bibinfo{journal}{Macromolecules} \textbf{\bibinfo{volume}{38}},
  \bibinfo{pages}{8078} (\bibinfo{year}{2005}).

\bibitem[{\citenamefont{Hess}(1986)}]{he86}
\bibinfo{author}{\bibfnamefont{W.}~\bibnamefont{Hess}},
  \bibinfo{journal}{Macromolecules} \textbf{\bibinfo{volume}{19}},
  \bibinfo{pages}{1395} (\bibinfo{year}{1986}).

\bibitem[{\citenamefont{Hess}(1988)}]{he88}
\bibinfo{author}{\bibfnamefont{W.}~\bibnamefont{Hess}},
  \bibinfo{journal}{Macromolecules} \textbf{\bibinfo{volume}{21}},
  \bibinfo{pages}{2620} (\bibinfo{year}{1988}).

\bibitem[{\citenamefont{Paul et~al.}(1991)\citenamefont{Paul, Binder, Heermann,
  and Kremer}}]{pa91}
\bibinfo{author}{\bibfnamefont{W.}~\bibnamefont{Paul}},
  \bibinfo{author}{\bibfnamefont{K.}~\bibnamefont{Binder}},
  \bibinfo{author}{\bibfnamefont{D.~W.} \bibnamefont{Heermann}},
  \bibnamefont{and} \bibinfo{author}{\bibfnamefont{K.}~\bibnamefont{Kremer}},
  \bibinfo{journal}{J. Phys. II} \textbf{\bibinfo{volume}{1}},
  \bibinfo{pages}{37} (\bibinfo{year}{1991}).

\bibitem[{\citenamefont{Shaffer}(1994)}]{sh94}
\bibinfo{author}{\bibfnamefont{J.~S.} \bibnamefont{Shaffer}},
  \bibinfo{journal}{J. Chem. Phys.} \textbf{\bibinfo{volume}{101}},
  \bibinfo{pages}{4205} (\bibinfo{year}{1994}).

\bibitem[{\citenamefont{Shaffer}(1995)}]{sh95}
\bibinfo{author}{\bibfnamefont{J.~S.} \bibnamefont{Shaffer}},
  \bibinfo{journal}{J. Chem. Phys.} \textbf{\bibinfo{volume}{103}},
  \bibinfo{pages}{761} (\bibinfo{year}{1995}).

\bibitem[{\citenamefont{Pan and Shaffer}(1996)}]{pa96}
\bibinfo{author}{\bibfnamefont{X.}~\bibnamefont{Pan}} \bibnamefont{and}
  \bibinfo{author}{\bibfnamefont{J.~S.} \bibnamefont{Shaffer}},
  \bibinfo{journal}{Macromolecules} \textbf{\bibinfo{volume}{29}},
  \bibinfo{pages}{4453} (\bibinfo{year}{1996}).

\bibitem[{\citenamefont{Newman and Barkema}(1999)}]{ne99}
\bibinfo{author}{\bibfnamefont{M.~E.~J.} \bibnamefont{Newman}}
  \bibnamefont{and} \bibinfo{author}{\bibfnamefont{G.~T.}
  \bibnamefont{Barkema}}, \emph{\bibinfo{title}{Monte Carlo methods in
  statistical physics}} (\bibinfo{publisher}{Clarendon Press},
  \bibinfo{address}{Oxford, UK}, \bibinfo{year}{1999}).

\bibitem[{\citenamefont{Frenkel and Smit}(1996)}]{fr96b}
\bibinfo{author}{\bibfnamefont{D.}~\bibnamefont{Frenkel}} \bibnamefont{and}
  \bibinfo{author}{\bibfnamefont{B.}~\bibnamefont{Smit}},
  \emph{\bibinfo{title}{Understanding Molecular Simulations - From Algorithms
  to Applications}} (\bibinfo{publisher}{Academic Press}, \bibinfo{address}{San
  Diego, CA}, \bibinfo{year}{1996}).

\bibitem[{\citenamefont{Paul}(2002)}]{pa02b}
\bibinfo{author}{\bibfnamefont{W.}~\bibnamefont{Paul}},
  \bibinfo{journal}{Chemical Physics} \textbf{\bibinfo{volume}{284}},
  \bibinfo{pages}{59} (\bibinfo{year}{2002}).

\bibitem[{\citenamefont{McCormick et~al.}(2005)\citenamefont{McCormick, Hall,
  and Khan}}]{mc05}
\bibinfo{author}{\bibfnamefont{J.~A.} \bibnamefont{McCormick}},
  \bibinfo{author}{\bibfnamefont{C.~K.} \bibnamefont{Hall}}, \bibnamefont{and}
  \bibinfo{author}{\bibfnamefont{S.~A.} \bibnamefont{Khan}},
  \bibinfo{journal}{J. Chem. Phys.} \textbf{\bibinfo{volume}{122}},
  \bibinfo{pages}{114902} (\bibinfo{year}{2005}).

\bibitem[{\citenamefont{Pearson et~al.}(1987)\citenamefont{Pearson, Ver~Strate,
  von Meerwall, and Schilling}}]{pe87}
\bibinfo{author}{\bibfnamefont{D.~S.} \bibnamefont{Pearson}},
  \bibinfo{author}{\bibfnamefont{G.}~\bibnamefont{Ver~Strate}},
  \bibinfo{author}{\bibfnamefont{E.}~\bibnamefont{von Meerwall}},
  \bibnamefont{and} \bibinfo{author}{\bibfnamefont{F.~C.}
  \bibnamefont{Schilling}}, \bibinfo{journal}{Macromolecules}
  \textbf{\bibinfo{volume}{20}}, \bibinfo{pages}{1133} (\bibinfo{year}{1987}).

\bibitem[{\citenamefont{Pearson et~al.}(1994)\citenamefont{Pearson, Fetters,
  Graessley, Ver~Strate, and von Meerwall}}]{pe94}
\bibinfo{author}{\bibfnamefont{D.~S.} \bibnamefont{Pearson}},
  \bibinfo{author}{\bibfnamefont{L.~J.} \bibnamefont{Fetters}},
  \bibinfo{author}{\bibfnamefont{W.~W.} \bibnamefont{Graessley}},
  \bibinfo{author}{\bibfnamefont{G.}~\bibnamefont{Ver~Strate}},
  \bibnamefont{and} \bibinfo{author}{\bibfnamefont{E.}~\bibnamefont{von
  Meerwall}}, \bibinfo{journal}{Macromolecules} \textbf{\bibinfo{volume}{27}},
  \bibinfo{pages}{711} (\bibinfo{year}{1994}).

\bibitem[{\citenamefont{Flory}(1953)}]{fl53}
\bibinfo{author}{\bibfnamefont{P.~J.} \bibnamefont{Flory}},
  \emph{\bibinfo{title}{Principles of Polymer Chemistry}}
  (\bibinfo{publisher}{Cornell University}, \bibinfo{address}{Ithaca, NY},
  \bibinfo{year}{1953}).

\bibitem[{\citenamefont{Dickman and Hall}(1986)}]{di86}
\bibinfo{author}{\bibfnamefont{R.}~\bibnamefont{Dickman}} \bibnamefont{and}
  \bibinfo{author}{\bibfnamefont{C.~K.} \bibnamefont{Hall}},
  \bibinfo{journal}{J. Chem. Phys.} \textbf{\bibinfo{volume}{85}},
  \bibinfo{pages}{3023} (\bibinfo{year}{1986}).

\bibitem[{\citenamefont{Dickman}(1987)}]{di87}
\bibinfo{author}{\bibfnamefont{R.}~\bibnamefont{Dickman}}, \bibinfo{journal}{J.
  Chem. Phys.} \textbf{\bibinfo{volume}{87}}, \bibinfo{pages}{2246}
  (\bibinfo{year}{1987}).

\bibitem[{\citenamefont{Inoue and Osaki}(1996)}]{in96}
\bibinfo{author}{\bibfnamefont{T.}~\bibnamefont{Inoue}} \bibnamefont{and}
  \bibinfo{author}{\bibfnamefont{K.}~\bibnamefont{Osaki}},
  \bibinfo{journal}{Macromolecules} \textbf{\bibinfo{volume}{29}},
  \bibinfo{pages}{1595} (\bibinfo{year}{1996}).

\bibitem[{\citenamefont{Luettmer-Strathmann and Khatri}()}]{lu06}
\bibinfo{author}{\bibfnamefont{J.}~\bibnamefont{Luettmer-Strathmann}}
  \bibnamefont{and} \bibinfo{author}{\bibfnamefont{R.}~\bibnamefont{Khatri}},
  \bibinfo{howpublished}{in preparation}.

\end{thebibliography}

\end{document}